\documentclass[prd,notitlepage,showpacs,nofootinbib,superscriptaddress]{revtex4-2}

\usepackage[utf8]{inputenc} 

\usepackage{amsmath}
\usepackage{amssymb}
\usepackage{float}
\usepackage{comment}
\usepackage{slashed}
\usepackage[normalem]{ulem}
\usepackage{bbold}
\usepackage{cancel}
\usepackage{graphicx}
\usepackage{multirow}
\usepackage{rotating}
\usepackage{caption}
\usepackage{subcaption}
\usepackage{tabulary}

\newcolumntype{K}[1]{>{\centering\arraybackslash}m{#1}}
\def\gsim{\raise0.3ex\hbox{$\;>$\kern-0.75em\raise-1.1ex\hbox{$\sim\;$}}}
\def\lsim{\raise0.3ex\hbox{$\;<$\kern-0.75em\raise-1.1ex\hbox{$\sim\;$}}}

\usepackage[usenames,dvipsnames]{color}

\newcommand {\ignore}[1]{}

\usepackage[colorinlistoftodos]{todonotes}
\usepackage[colorlinks=true,linkcolor=darkblue,citecolor=darkblue,urlcolor=darkblue, pdfborder={0 0 0}]{hyperref}
\usepackage[normalem]{ulem}

\usepackage{caption}
\usepackage{subcaption}
\definecolor{linkcolor}{rgb}{0,0,0.8}

\definecolor{darkgreen}{rgb}{0,0.5,0}
\definecolor{darkred}{rgb}{0.6,0,0}
\definecolor{brown}{rgb}{0.59, 0.29, 0.0}
\definecolor{mightnightblue}{RGB}{25,25,112}
\definecolor{darkblue}{rgb}{0,0,0.8}

\newcommand {\darkblue} {\color{darkblue}}

\def\EW{$\mathrm{SU(2)_L \otimes U(1)_Y}$ }

\newcommand{\M}{\mathbf{M}}
\newcommand{\U}{\mathbf{U}}

\newcommand{\BR}{{\rm BR}}

\def\Mnu{\mathbf{M}_\nu}

\def\U{\mathbf{U}}

\newcommand{\AddrCFTP}{%
Departamento de F\'{\i}sica and CFTP\\ Instituto Superior T\'ecnico, Universidade de Lisboa\\ Av. Rovisco Pais, 1\\ 1049-001 Lisboa, Portugal}

\usepackage[force]{feynmp-auto}

\begin{document}

\title{\darkblue \Large  Dark linear seesaw mechanism}

\author{\bf A. Batra}\email{aditya.batra@tecnico.ulisboa.pt}
\affiliation{\AddrCFTP}

\author{\bf H.~B. C\^amara}\email{henrique.b.camara@tecnico.ulisboa.pt}
\affiliation{\AddrCFTP}

\author{\bf F.~R. Joaquim}\email{filipe.joaquim@tecnico.ulisboa.pt }
\affiliation{\AddrCFTP}

\begin{abstract}
\vspace{0.2cm}
\begin{center}
{ \center  \bf ABSTRACT}\\    
\end{center}
We propose a minimal model where a \emph{dark} sector, odd under a $\mathcal{Z}_2$ discrete symmetry, is the seed of lepton number violation in the neutrino sector at the loop level, in the context of the linear seesaw mechanism. Neutrino mass suppression stems from a naturally small scalar potential coupling which breaks the lepton number symmetry softly. The fact that we consider (\emph{dark}) Dirac vector-like neutral leptons, prevents the appearance of other mass terms that could contribute to neutrino masses via alternative mechanisms. We study the dark-matter phenomenology of the model, focusing on the case in which the stable particle is the lightest neutral scalar arising from the \emph{dark} scalar sector. Prospects for testing our framework with the results of current and future lepton flavour violation searches are also discussed.
\end{abstract}

\maketitle
\noindent

\section{Introduction}
\label{sec:intro}

The discovery of neutrino flavour oscillations~\cite{Kajita:2016cak,McDonald:2016ixn} and the fact that about $27 \%$ of the total matter in the Universe appears in the form of dark matter (DM)~\cite{Bertone:2004pz,Planck:2018vyg} provide two clear evidences for physics beyond the Standard Model (SM). From a theory perspective, addressing these issues requires extending the SM symmetry and its field content. Several mechanisms have been proposed to generate naturally small neutrino masses, either at tree level~\cite{Minkowski:1977sc,Gell-Mann:1979vob,Yanagida:1979as,Schechter:1980gr,Glashow:1979nm,Mohapatra:1979ia}, or radiatively~\cite{Zee:1980ai,Ma:1998dn}. In the former class of models, the type-I seesaw~\cite{Minkowski:1977sc,Gell-Mann:1979vob,Yanagida:1979as,Schechter:1980gr,Glashow:1979nm,Mohapatra:1979ia} requires extremely heavy right-handed neutrinos with masses well above the electroweak~(EW) scale, suppressing effective (Majorana) neutrino masses. In contrast, low-scale variants such as the inverse~\cite{Mohapatra:1986bd} and linear~\cite{Akhmedov:1995ip,Akhmedov:1995vm,Malinsky:2005bi} seesaws, employ schemes where heavy neutrino masses can be lowered down to the TeV scale. This enhances the prospects of probing and testing the seesaw paradigm at current and future experiments by looking at, for instance, charged lepton flavour violation~(cLFV)~\cite{Alonso:2012ji,Abada:2015oba,Abada:2018nio,Hernandez-Tome:2019lkb,Camara:2020efq}. In radiative models, small neutrino masses are induced at the quantum level and, interestingly, the particles entering the loops may also be viable DM candidates stabilized by some symmetry as a simple $\mathcal{Z}_2$. This is the principle behind the scotogenic mechanism~\cite{Ma:2006km} that combines neutrino mass generation and DM in a framework where DM acts as the ``seed'' of neutrino masses (for a review on radiative neutrino mass models see Ref.~\cite{Cai:2017jrq}).

Inspired in these ideas, we propose a simple model where DM ``seeds'' lepton-number breaking in the linear-seesaw context with a minimal field and symmetry content. Namely, assuming a (softly-broken) lepton-number symmetry U$(1)_L$, we add three neutral fermion singlets with $L=\pm 1$. DM stability is ensured by a $\mathcal{Z}_2$ discrete symmetry under which one of the fermions is odd. The soft U$(1)_L$ breaking stems from a cubic interaction among the SM Higgs doublet and two \emph{dark} scalars (a doublet and a real singlet). This Letter is organised as follows. In Sec.~\ref{sec:model}, we present the simplest model where DM generates radiatively the linear seesaw based on explicit global lepton number breaking. The phenomenological implications of the model in regards to DM and cLFV are studied in Sec.~\ref{sec:pheno}. Finally, we state our concluding remarks in Sec.~\ref{sec:concl}. Details about the scalar sector and the generalised Casas-Ibarra~(CI) parameterisation are shown in the appendices.

\section{The dark linear seesaw}
\label{sec:model}

%
\begin{table}[t!]
	\centering
	\begin{tabular}{| K{1.5cm} | K{1cm} | K{2.5cm} | K{1.0cm} | K{1.0cm} | }
		\hline 
&Fields&\EW&  U$(1)_L$ &  $\mathcal{Z}_2$  \\
		\hline 
		\multirow{4}{*}{Fermions} 
&$L$&($\mathbf{2}, {-1}$)& $1$   &   $+$  \\
&$e_R$&($\mathbf{1}, {2}$)& {$1$}  &   $+$   \\
&$\nu_R$&($\mathbf{1}, {0}$)& {$1$}  &  $+$   \\
&$S_R$&($\mathbf{1}, {0}$)& {$-1$} &    $+$\\
&$f_{L,R}$&($\mathbf{1}, {0}$)& {$-1$}   & {$-$} \\
		\hline 
\multirow{3}{*}{Scalars}  &$\Phi$&($\mathbf{2}, {1}$)& {$0$} &  $+$   \\
&$\eta$&($\mathbf{2}, {1}$)& {$-2$}  &  $-$   \\
&$\chi$&($\mathbf{1},0$)& {$0$} &  $-$   \\		
\hline
	\end{tabular}
	\caption{Field content and transformation properties under \EW and U$(1)_L \otimes \mathcal{Z}_2$.}
	\label{tab:model} 
\end{table}
The linear seesaw mechanism is implemented by adding sterile singlet fermions $\nu_R$ and $S_R$ to the SM, requiring in its minimal realisation one of each to explain the two observed neutrino mass squared differences. In this framework, lepton number is violated by the $m_S\overline{\nu_L}S_R$ term, where $\nu_L$ is the SM neutrino field from the lepton doublet $\ell_L$. Neutrino masses are then proportional to $m_S$ which can be naturally small in the t'Hooft sense~\cite{tHooft:1979rat} and, hence, be the origin of neutrino mass suppression. Due to the existence of that small lepton number violating (LNV) parameter, low-scale neutrino mass generation can be envisaged, in contrast with the type-I seesaw which requires extremely heavy mediators. 

In this work, we consider the possibility that LNV in the neutrino sector is seeded at the quantum level by \emph{dark} fields, allowing for a direct connection to the DM problem. Our setup is based on the symmetry and field content shown in Table~\ref{tab:model}, from which it is apparent that the global lepton number symmetry U$(1)_L$ forces $m_S=0$ at tree-level. To generate it radiatively, we introduce a Dirac vector-like singlet fermion $f_{L,R}$, as well as a scalar doublet $\eta$ and real singlet $\chi$. These three fields (the \emph{dark}-sector) are odd under a discrete $\mathcal{Z}_2$ symmetry. The most general fermion mass and Yukawa Lagrangian allowed by the SM gauge and U$(1)_L \otimes \mathcal{Z}_2$ symmetries is:
\begin{align}
    -\mathcal{L} = \mathbf{Y}_e \overline{\ell_L} \Phi e_R + \mathbf{Y}_D \overline{\ell_L} \tilde{\Phi} \nu_R + \mathbf{Y}_f \overline{\ell_L} \tilde{\eta} f_R + Y_S \overline{f_L} S_R \chi + Y_{R} \overline{f_R^c} \nu_R \chi + M_B \overline{\nu_R^c} S_R + M_f \overline{f_L} f_R + \text{H.c.} \; ,
    \label{eq:LYuk}
\end{align}
where the Yukawa couplings $\mathbf{Y}_e$, $\mathbf{Y}_{D,f}$ and $Y_{R,S}$ are $3 \times 3$ matrices, $3 \times 1$ vectors and numbers, respectively. The mass $M_B$ is the typical heavy-neutrino mass scale, while $M_f$ is the bare mass of the \emph{dark} fermion. Without loss of generality, we consider that the charged leptons are already in the physical basis and focus on the neutrino sector only. The scalar doublets are defined as $\Phi = (\phi^+ , \phi^0)^T$ and $\eta = (\eta^+ , \eta^0)^T$ with $\tilde{\Phi}= i \tau_2 \Phi^\ast$ and $\tilde{\eta}= i \tau_2 \eta^\ast$ ($\tau_2$ is the complex Pauli matrix). The vacuum configuration of the model is such that $\left<\phi^0\right> = v/\sqrt{2}\simeq 174$~GeV and, for the $\mathcal{Z}_2$ DM symmetry to remain unbroken, $\eta$ and $\chi$ must be VEVless. We will see in Sec.~\ref{sec:DM} that either the lightest \emph{dark} scalar or fermion $f$ can be suitable weakly interacting massive particle~(WIMP) DM. After EW symmetry breaking, the full tree-level neutrino mass matrix $\mathcal{M}_\nu^0$ defined in the $(\nu_L, \nu_R^c,S_R^c)$ basis is,
\begin{align}
\mathcal{M}_\nu^0 = \begin{pmatrix}
 0 & \mathbf{M}_D & 0    \\
 \mathbf{M}_D^{T} & 0 & M_B    \\
 0 & M_B & 0    \\
\end{pmatrix} \; , \; \mathbf{M}_D = \frac{v \mathbf{Y}_D}{\sqrt{2}} \; .
\label{eq:treemassmatrix}
\end{align}
Obviously, active neutrinos are massless at tree level due to lepton number conservation.
    \begin{figure}[t!]
        \centering
        \includegraphics[scale=0.9]{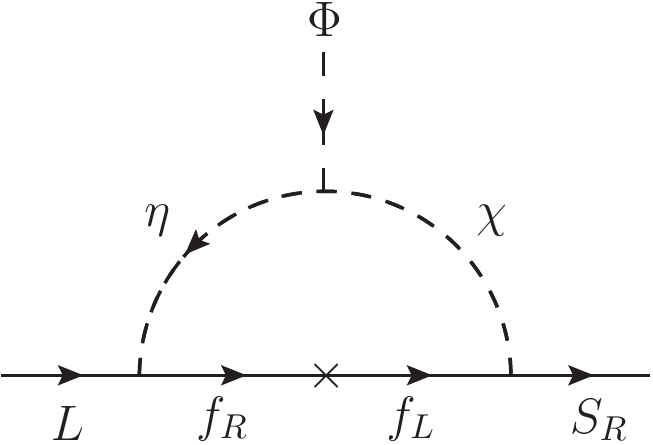} \hspace{+1cm} \includegraphics[scale=0.9]{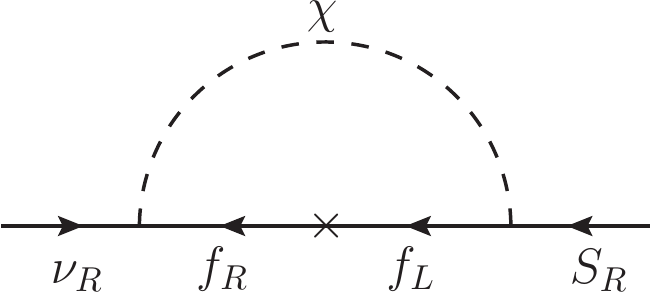}
        \caption{Lowest order one-loop diagrams for neutrino mass generation: \emph{dark}-seeded linear seesaw (left) and additional correction to the bare heavy neutrino mass (right) (see text for details).}
    \label{fig:neutrino}
    \end{figure}

The origin of lepton number violation stems from the scalar potential term:
    \begin{align}
    V_{\text{soft}} = \kappa \left(\eta^\dagger \Phi \right)\chi + \text{H.c.} \; ,
    \label{eq:Vpotential}
    \end{align}
which breaks U$(1)_L$ softly by two units. Note that, this is the only possible soft LNV term that can be written in the scalar potential. This is due to the restriction imposed by the $\mathcal{Z}_2$ symmetry which forbids any other soft LNV term that might lead to the decay of \emph{dark} sector particles. This explicit lepton number breaking triggers radiative neutrino mass generation with the $\mathcal{Z}_2$ odd fields $\eta$, $\chi$ and $f_{L,R}$ running in the loops, as shown in Fig.~\ref{fig:neutrino}. Namely, the left diagram is responsible for inducing a nonzero $m_S$, whose smallness comes from the naturally-small LNV parameter $\kappa$ and \emph{dark}-loop suppression. Furthermore, the diagram on the right corresponds to a \emph{dark} radiative correction $\Delta M_B$ to the bare mass term $M_B$. Note that the \emph{dark} fermion $f_{L,R}$ being a Dirac particle prevents the existence of any higher order corrections to Majorana mass terms for $\nu_L$, $\nu_R$ or $S_R$ (the diagonal entries in the neutrino mass matrix). Taking all this into account, the neutrino mass matrix can be now written as
\begin{align}
\mathcal{M}_\nu = \mathcal{M}_\nu^0 + \Delta \mathcal{M}_\nu = \begin{pmatrix}
 0 & \mathbf{M}_D & \mathbf{m}_S    \\
 \mathbf{M}_D^T & 0 & M    \\
 \mathbf{m}_S^T & M & 0    \\
\end{pmatrix} \; , \; M = M_B + \Delta M_B \; .
\end{align}
where $\mathbf{m}_S$ (a $3\times 1$ vector) is the equivalent of the generic tree-level $m_S$ in the regular linear seesaw. The one-loop mass terms computed in the scalar mass basis can be written as
\begin{align}
 \mathbf{m}_S = \mathcal{F}_{S}(M_f,m_{\zeta_k}) M_f \; Y_S \mathbf{Y}_f \;, \; 
 \Delta M_B  = \mathcal{F}_{B}(M_f,m_{\zeta_k}) M_f \; Y_{R} Y_S \; ,
\label{eq:loops}
\end{align}
where the loop factors $\mathcal{F}_{S,B}(M_f,m_{\zeta_k})$ depend on $M_f$, the \textit{dark} neutral-scalar masses $m_{\zeta_k}$ ($k=1, 2, 3$) and the mixing matrix of the neutral components of $\eta$ and $\chi$. The complete expressions for $\mathcal{F}_{S,B}(M_f,m_{\zeta_k})$ are given in  Eq.~\eqref{eq:Floop} of Appendix~\ref{sec:scalar}, where other details regarding the scalar potential and mass spectrum can also be found. 

In the linear seesaw approximation, i.e. for $\mathbf{m}_S \ll \mathbf{M}_D \ll M$,  $\mathcal{M}_\nu $ can be diagonalised as in Ref.~\cite{Schechter:1981cv}, leading to the effective light neutrino mass matrix,
\begin{align}
\mathbf{M}_\nu \simeq - \frac{\mathbf{M}_D \mathbf{m}_S^T + \mathbf{m}_S \mathbf{M}_D^T}{M} \;.
\label{eq:Mnu}
\end{align}
Note that, since we consider the minimal setup containing a single $\nu_R-S_R$ pair, $\mathbf{M}_\nu$ has rank $2$. Therefore, this scenario provides a massless neutrino at one-loop level, while with two $\nu_R-S_R$ all three light neutrinos would be massive. The small parameter $\boldsymbol{\epsilon}$ relevant for the seesaw diagonalisation procedure, in the limit $\mathbf{m}_S \rightarrow 0$, is
\begin{align}
\boldsymbol{\epsilon} \simeq - \frac{\mathbf{M}_D}{M} \; \Rightarrow \; \mathbf{K}=\begin{pmatrix}
 \mathbf{K}_{\text{light}}  &  \mathbf{K}_{\text{heavy}} 
 \end{pmatrix} \; , \; \mathbf{K}_{\text{light}} =\left(1-\frac{1}{2} \boldsymbol{\epsilon}^\ast \boldsymbol{\epsilon}^T\right) \mathbf{U}_{\nu} \; , \; \mathbf{K}_{\text{heavy}} =\left(\frac{\boldsymbol{\epsilon}^\ast}{\sqrt{2}}S  \,\,\,\,\, \frac{i \boldsymbol{\epsilon}^\ast}{\sqrt{2}} \right) \; ,
 \label{eq:K} 
\end{align}
where $\mathbf{K}$ is the $3 \times 5$ rectangular matrix which appears in the charged-current interactions and induces cLFV as discussed in Sec.~\ref{sec:cLFV}. $\mathbf{K}_{\text{light}}$ is the $3 \times 3$ matrix that diagonalises the effective light neutrino mass matrix with $\boldsymbol{\epsilon}^\ast \boldsymbol{\epsilon}^T$ encoding deviations from unitarity of the lepton mixing matrix $\mathbf{U}_{\nu}$. $\mathbf{K}_{\text{heavy}}$ is the $3 \times 2$ heavy-light mixing matrix. Neglecting those deviations, the effective neutrino mass matrix can be diagonalised through a unitary rotation $\nu \rightarrow \mathbf{U}_{\nu}\, \nu$:
\begin{equation}
\mathbf{U}_{\nu}^{ T}\, \Mnu\, \mathbf{U}_{\nu}  = \mathbf{d}_{\nu} = \text{diag}\left(m_1, m_2, m_3\right) \; , \; m_4 \simeq m_5 \simeq M \; ,
\label{eq:Unudef}
\end{equation}
where $m_{1,2,3}$ are the real and positive light neutrino masses. The heavy neutrinos stemming from $\nu_R$ and $S_R$ form a pseudo-Dirac pair, with masses $m_{4,5}$ approximately equal to $M$ apart from terms proportional to the small parameters $\mathbf{m}_S$ and $\boldsymbol{\epsilon}$. Note that, as mentioned before, the linear seesaw can be accommodated as a low-scale mechanism. In fact, since the smallness of $\mathbf{m}_S$ is provided by the LNV parameter $\kappa$ and loop suppression, we can have $\mathbf{m}_S \sim \mathcal{O}(1\,{\rm eV})$, which leads to light neutrino masses $\sim \mathcal{O}(0.1\,{\rm eV})$ for a heavy neutrino mass scale at $M \sim \mathcal{O}(1\,{\rm TeV})$. This will have important phenomenological implications in regards to cLFV~(see Sec.~\ref{sec:cLFV}). The unitary matrix $\mathbf{U}_{\nu}$ above is the lepton mixing matrix which can be parameterised by three mixing angles $\theta_{12}$, $\theta_{23}$, and $\theta_{13}$, and two CP-violating phases: a Dirac-type phase $\delta$ and a single Majorana-type phase $\alpha$ (since we have a massless neutrino)~\cite{ParticleDataGroup:2020ssz}. Besides the mixing angles and Dirac phase $\delta$, neutrino oscillation experiments constrain the two neutrino mass squared differences $\Delta m_{21}^2= m_2^2 - m_1^2$ and $\Delta m_{31}^2= m_3^2 - m_1^2$, where there are two possible orderings for neutrino masses: normal and inverted ordering (NO and IO, respectively). Since in our minimal case the lightest neutrino is massless, we have $m_1=0$ for NO and $m_3=0$ for IO. The framework presented here is testable at future experiments looking for the neutrinoless double beta decay process which could elucidade on the nature of neutrinos, i.e. if they are of Majorana or Dirac type~\cite{Schechter:1981bd}. In fact, in the case of a massless neutrino, the effective Majorana mass parameter~\cite{Barreiros:2018bju} has a lower bound for NO neutrino masses while, most importantly, the IO prediction falls within the sensitivity of future experiments such as SNO+ II~\cite{SNO:2015wyx}, LEGEND~\cite{LEGEND:2017cdu} and nEXO~\cite{nEXO:2017nam}.

\section{Phenomenology}
\label{sec:pheno}

%
\begin{table}[!t]
\centering
\begin{tabular}{|K{1.3cm}|K{2cm}|K{3cm}|}   
\hline
Sector & Parameters & Scan range \\
\hline
\multirow{3}{*}{Fermion} & $M_f$ & $[10 , 10^4]$ (GeV)  \\
& $M_B$ & $[10 , 10^5]$ (GeV) \\
& $Y_S, Y_{R}$ & $\left[10^{-8} , 1\right]$ \\
\hline
\multirow{3}{*}{Scalar} & $m_{\eta}^2 , m_{\chi}^2$ &  $[10^2 , 10^8]$ (GeV$^2$) \\
& $\kappa$ &  $[10^{-8}, 10^{2}]$ (GeV)  \\
& $\lambda_{3},|\lambda_{4}|,\lambda_{\Phi \chi}$ & $[10^{-8} , 1]$ \\
\hline
\end{tabular}
\caption{Input parameters of our model and corresponding ranges used in our numerical scan (see text for details). }
\label{tab:Scan}
\end{table}
In this section, we turn our attention to the DM and cLFV phenomenological implications of our model. We perform a numerical scan with scalar and fermion input parameters [see Eqs.~\eqref{eq:LYuk} and~\eqref{eq:Vpotentialfull}, respectively] as shown in Table~\ref{tab:Scan}, where for simplicity we assume that $Y_{R,S}$ are real. We set the SM Higgs mass to $m_h=125.09$ GeV~\cite{ParticleDataGroup:2020ssz} and \emph{dark}-scalar self-couplings to $\lambda_{\eta} = \lambda_{\chi} = \lambda_{\eta \chi} = 0.5$ since they do not modify our analysis. The scalar sector is analysed in detail in Appendix~\ref{sec:scalar}, where scalar masses and mixing are given, as well as the vacuum stability conditions which we took into account in our numerical analysis. Namely, by requiring the scalar potential to be bounded from below [see Eq.~\eqref{eq:bounded}] and the scalar masses to be positive are sufficient conditions to ensure that the VEV configuration of the scalars is the global minimum of the potential. To reconstruct the Yukawa vectors $\mathbf{Y}_{D,f}$, we use the generalised CI parameterisation given in Appendix~\ref{sec:CI} with a real angle $z \in [0,2\pi]$, and $\alpha \in [0,2\pi]$ for the Majorana phase. Note that, we focus on the currently favoured NO neutrino masses and vary the values for the neutrino oscillation observables (neutrino mass squared-differences, mixing angles and Dirac CP phase) within their $3 \sigma$ ranges using the results obtained by the global data fit of Ref.~\cite{deSalas:2020pgw}.

\subsection{Dark matter}
\label{sec:DM}

The presence of the $\mathcal{Z}_2$ symmetry ensures the stability of the lightest $\mathcal{Z}_2$ odd particle. Thus, the fermion $f$ and the scalar $\zeta_1$ (which is a mixed state of the inert scalar doublet $\eta$ and singlet $\chi$) can be viable WIMP DM candidates. Here, we focus on the scalar DM phenomenology, i.e. when the lightest \emph{dark} particle is $\zeta_1$. Hence, the $\zeta_1-\zeta_1$ annihilation channels and coannihilations $\zeta_1- \zeta_k$ ($k= 2, 3$), $\zeta_1-\zeta^{\pm}$ and $\zeta_1-f$, into $\mathcal{Z}_2$ even particles will contribute to the thermaly-averaged cross section and consequently to the DM relic density. The current observed value for DM relic abundance obtained by the Planck satellite data, at the $3 \sigma$ level, is~\cite{Planck:2018vyg},
    \begin{equation}
   0.1126 \leq \Omega_{\text{DM}} h^2 \leq 0.1246 \; ,
   \label{eq:Oh2Planck}
    \end{equation}
where $h$ is the reduced Hubble constant. To check whether $\zeta_1$ is a viable DM candidate, we implement our model in \texttt{SARAH}~\cite{Staub:2013tta} and use \texttt{SPheno}~\cite{Porod:2003um} to obtain mass matrices, vertices and tadpole equations. The relic density $\Omega h^2$ and WIMP-nucleon spin-independent elastic scattering cross-section $\sigma^{\text{SI}}$ are computed at tree-level by the \texttt{micrOMEGAS}~\cite{Belanger:2014vza} package. In our numerical scan (see Table~\ref{tab:Scan}), we apply the following experimental constraints: 
\begin{itemize}
    
    \item \textit{LEP:} The precise LEP-I measurements on the $Z$-boson decay width leads to the lower bounds on the \textit{dark}-neutral scalar masses, $m_{\zeta_i} > m_Z/2 = 45.6$ GeV and $m_{\zeta_i} + m_{\zeta_j} > m_Z = 91.2$ GeV ($i , j = 1, 2, 3$)~\cite{Cao:2007rm,Gustafsson:2007pc}. The latter ensures that the decays $Z \rightarrow \zeta_i \zeta_k$, are kinematically forbidden. Also, reinterpreting LEP-II results for chargino searches in the context of singly-charged scalar production, leads to the conservative bound on the \textit{dark}-charged scalar mass $ m_{\zeta^+} > 70$ GeV~\cite{Pierce:2007ut}.

     \item \textit{LHC Higgs data}: The odd-neutral scalars $\zeta_i$ ($i= 1, 2, 3$) contribute to the Higgs invisible decay width through the channels $h \rightarrow \zeta_i \zeta_j$ ($i,j= 1, 2, 3$), which open up for masses $m_{\zeta_i} + m_{\zeta_j} < m_{h}=125.09$ GeV. The branching ratio~(BR) for this process, is constrained by the LHC Higgs data~\cite{ATLAS:2016neq,ATLAS:2019nkf,CMS:2018yfx}, with the current bound being $\BR(h \rightarrow \text{inv}) \leq 0.19$~\cite{ParticleDataGroup:2020ssz}. The odd-charged scalar $\zeta^+$ will contribute at loop level to the diphoton Higgs decay $h \rightarrow \gamma \gamma$ with a signal strength such that~\cite{ATLAS:2018hxb}
     \begin{equation}
     R_{\gamma\gamma}=\frac{\text{BR}(h \to \gamma\gamma)}{\text{BR}(h \to \gamma\gamma)_{\text{SM}}}=0.99^{+0.15}_{-0.14} \,.
     \end{equation}

    \item \textit{cLFV:} The BR of the $\mu \to e \gamma$ process is constrained by MEG to be less than $4.2 \times 10^{-13}$~\cite{MEG:2016leq}. This is discussed in further detail in Sec.~\ref{sec:cLFV}.
    
\end{itemize}
\begin{figure}[!t]
    \centering
    \includegraphics[scale=0.34]{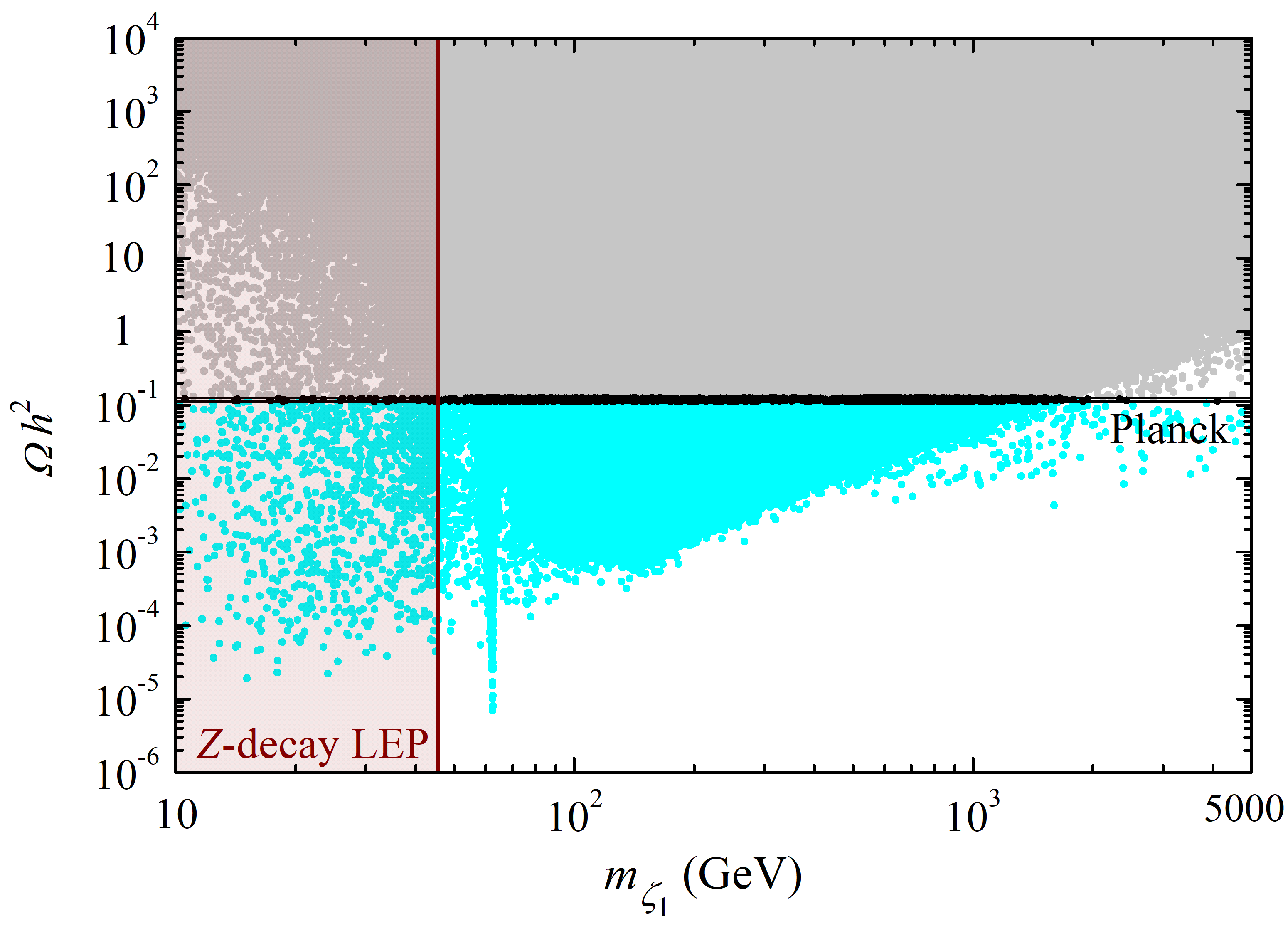}\includegraphics[scale=0.34]{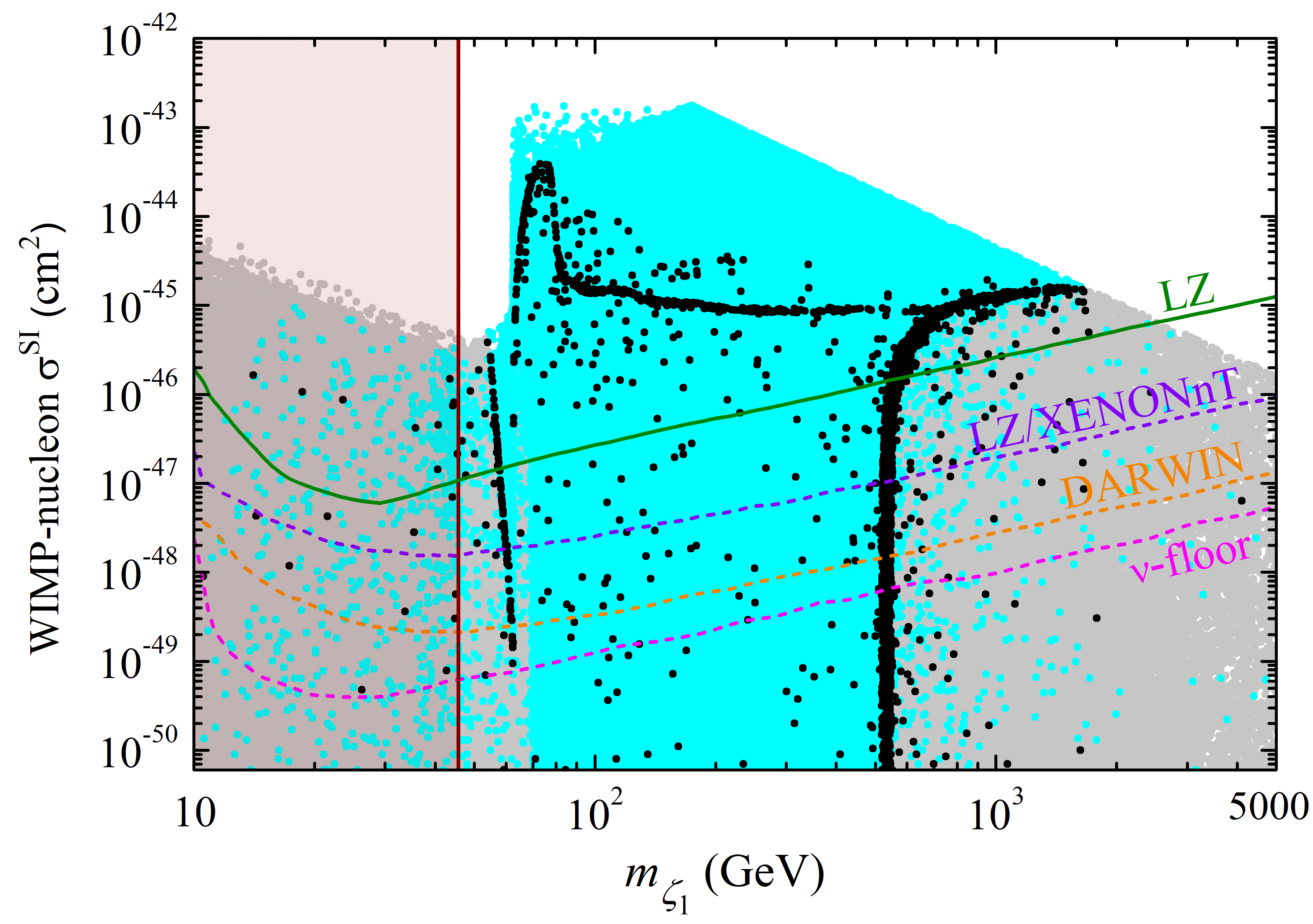}
    \caption{Relic density $\Omega h^2$ (left) and WIMP-nucleon spin-independent elastic scattering cross-section $\sigma^{\text{SI}}$ (right) versus the scalar DM mass $m_{\zeta_1}$. Black points lie within the $3 \sigma$ range of the measured relic abundance obtained by Planck~\cite{Planck:2018vyg} [see Eq.~\eqref{eq:Oh2Planck}]. The grey (cyan) points depict overabundant (underabundant) DM. The red-shaded region is excluded by LEP constraints (see text for details). On the right, the solid green line indicates the current bound from the DD experiment LZ~\cite{LZ:2022ufs}. The violet and orange-dashed contours indicate the projected sensitivities for LZ~\cite{LZ:2018qzl}, XENONnT~\cite{XENON:2020kmp}, and DARWIN~\cite{DARWIN:2016hyl}, respectively. The pink-dashed line is the "neutrino floor"~\cite{Billard:2013qya}.}
    \label{fig:DM}
\end{figure}
The results of our numerical analysis are presented in Fig.~\ref{fig:DM}. The left panel shows $\Omega h^2$ as a function of the scalar DM mass $m_{\zeta_1}$. The overabundant DM points shown in gray are excluded, while underabundance shown in cyan is still viable. The black points are within the 3$\sigma$ interval for the DM relic density obtained by Planck~\cite{Planck:2018vyg} [see Eq.~\eqref{eq:Oh2Planck}]. We remark that there is a dip at $m_{\zeta_1} \sim m_h/2$ due to DM annihilation mediated by the SM Higgs which becomes very efficient when it is on-shell. Note that, this dip is the only viable DM mass where correct relic abundance is achieved in the pure scalar singlet DM scenario~\cite{McDonald:1993ex,Guo:2010hq,Cline:2013gha,Feng:2014vea,Wu:2016mbe,GAMBIT:2017gge,Casas:2017jjg}. Furthermore, for masses $m_{\zeta_1} \gsim m_W, m_Z$, the annihilation processes into a pair of gauge bosons opens up leading to another dip in $\Omega h^2$. The most interesting feature of the model stems from the mixing between the neutral components of the \textit{dark} scalars $\eta$ and $\chi$. Namely, when DM is dominated by the doublet $\eta$, the relic density behaviour is similar to that in the inert/scotogenic model as shown in the lower thick band in the figure~\cite{Belyaev:2016lok,Mandal:2021yph,Avila:2019hhv}. However, the presence of the singlet $\chi$ via mixing with the doublet $\eta$ causes the existence of several points above this curve and consequently we have correct relic density points for a large mass interval $45 \ \text{GeV} \lsim m_{\zeta_1} \lsim 2 \ \text{TeV}$.

In the right panel, we show the WIMP-nucleon $\sigma^{\text{SI}}$ as a function of $m_{\zeta_1}$. Note that $\zeta_1$ will contribute to $\sigma^{\text{SI}}$ via tree-level diagrams mediated by the SM Higgs boson $h$ and the $Z$-boson, being the dominant contribution the one mediated by $h$. Also, there are several points between $45 \ \text{GeV} \lsim m_{\zeta_1} \lsim 2 \ \text{TeV}$ with the correct relic density that satisfy the current direct detection~(DD) bound imposed by the LZ experiment~\cite{LZ:2022ufs} (solid green line). Most of the points that evade the bound lie around $m_{\zeta_1} \sim 500$ GeV as in the canonical scotogenic model. Furthermore, a large part of the parameter space will be probed by future experiments such as LZ~\cite{LZ:2018qzl} and XENONnT~\cite{XENON:2020kmp} (purple-dashed line), and DARWIN~\cite{DARWIN:2016hyl} (orange-dashed contour), up to the "neutrino floor" (pink-dashed contour) coming from coherent elastic neutrino scattering~\cite{Billard:2013qya}.

Before closing this section we wish to comment on the alternative fermion DM, i.e. the case where the lightest \emph{dark} particle is $f$. This scenario would be very similar to what happens in scotogenic-type models, where for $M_f \gsim 45$ GeV the correct value for the relic density [see Eq.~\eqref{eq:Oh2Planck}] can be achieved as long as coannihilation channels between $f$ and $\zeta,\zeta^\pm$ are taken into account (these are important if the relative \emph{dark} fermion-scalar mass difference is below $\sim 10 \%$~\cite{Mahanta:2019gfe,Ahriche:2016cio,Hagedorn:2018spx,Barreiros:2022aqu}).

\subsection{Charged lepton flavour violation}
\label{sec:cLFV}

%
\begin{figure}[!t]
   \centering
   \includegraphics[scale=0.33]{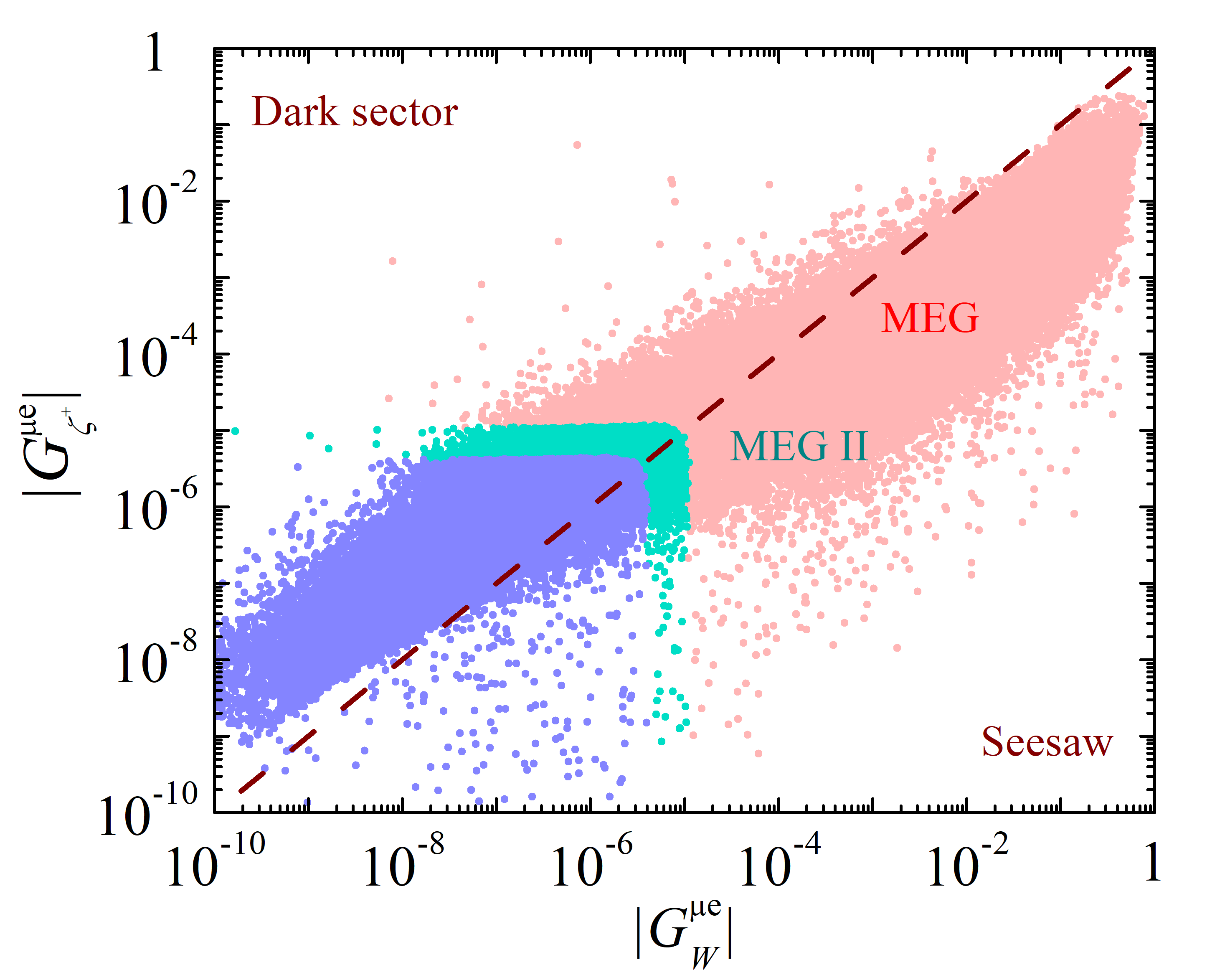} \\ 
  \vspace{+0.3cm} \includegraphics[scale=0.33]{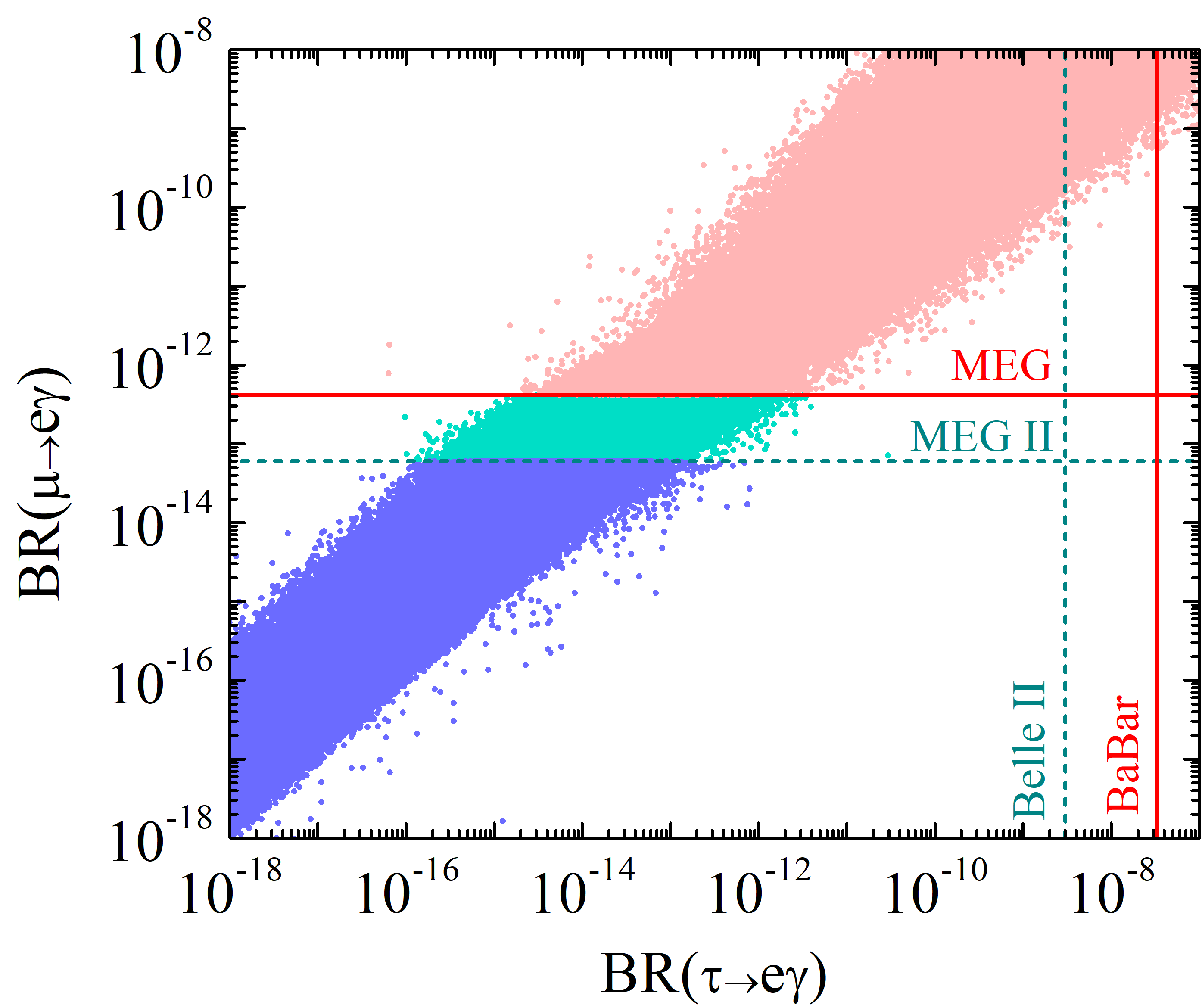} \hspace{+0.2cm} \includegraphics[scale=0.33]{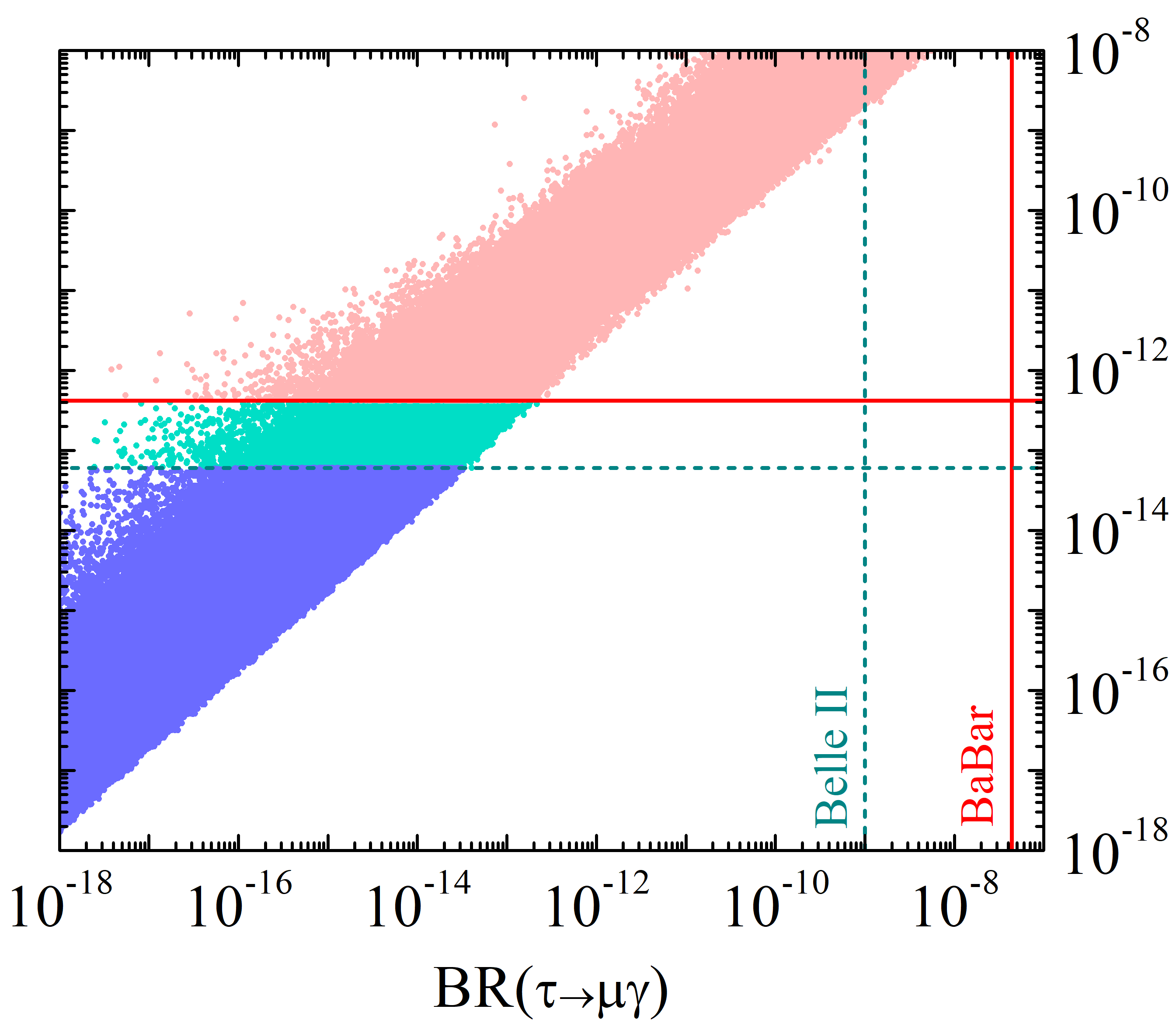}
  \caption{Top: $|G^{\mu e}_{\zeta^+}|$ versus $|G^{\mu e}_{W}|$ [see Eq.~\eqref{eq:cLFV} and discussion therein]. The red points are excluded by MEG~\cite{MEG:2016leq} while the green ones can be probed by MEG II~\cite{MEGII:2018kmf}. Along the brown-dashed contour $|G^{\mu e}_{\zeta^+}|=|G^{\mu e}_{W}|$ separating the \emph{dark}-sector from the seesaw cLFV dominance regime. Bottom: $\BR(\mu \rightarrow e \gamma)$ versus $\BR(\tau \rightarrow e \gamma)$ (left) and $\BR(\tau \rightarrow \mu \gamma)$ (right). Current cLFV bounds and projected sensitivities are marked by a red-solid and green-dashed lines, respectively.}
  \label{fig:cLFV}
\end{figure}
Searches for lepton flavour violating processes like $\ell_{\alpha} \rightarrow \ell_{\beta} \gamma$, $\ell_{\alpha} \rightarrow 3 \ell_{\beta}$ and $\mu - e$ conversion in nuclei are important probes of physics beyond the SM. These become even more relevant in low-scale seesaw scenarios since heavy-light neutrino mixing can be sizeable and induce non-negligible contributions to cLFV~\cite{Minkowski:1977sc,Marciano:1977wx,Cheng:1980tp,Lim:1981kv,Langacker:1988up,Ilakovac:1994kj,Alonso:2012ji,Abada:2015oba,Abada:2018nio,Hernandez-Tome:2019lkb,Camara:2020efq}. In our framework, there are additional loop contributions mediated by the \textit{dark}-sector particles $\zeta^\pm$ and $f$~\cite{Toma:2013zsa,Vicente:2014wga,Ahriche:2016cio,Hagedorn:2018spx,Barreiros:2022aqu}. Here, we will focus on radiative $\ell_{\alpha} \rightarrow \ell_{\beta} \gamma$ decays for which the branching ratios (BRs) are:
\begin{align}
\frac{\BR(\ell_{\alpha} \rightarrow \ell_{\beta} \gamma)}{\BR\left(\ell_{\alpha} \rightarrow \ell_{\beta} \nu_{\alpha} \overline{\nu_{\beta}} \right)} & = \frac{3 \alpha_{e}}{2 \pi} \left| G_W^{\alpha \beta}+ G_{\zeta^{+}}^{\alpha \beta}\right|^2 = \frac{3 \alpha_{e}}{2 \pi} \left|\sum_{i=1}^5 \mathbf{K}_{\alpha i}^{*} \mathbf{K}_{\beta i} G_W\left(\frac{m_{i}^2}{M_W^2}\right)+\frac{v^2}{4 m_{\zeta^\pm}^2} \mathbf{Y}_{f}^{\beta} \mathbf{Y}_{f}^{\alpha \ast}  G_{\zeta^{+}}\left(\frac{M_f^2}{m_{\zeta^{+}}^2}\right)\right|^2 \; ,
\label{eq:cLFV}
\end{align}
where $\alpha_{e} = e^2/(4\pi)$, $\BR\left(\mu \rightarrow e \nu_{\mu} \overline{\nu_{e}}\right) \simeq 1.0$, $\BR\left(\tau \rightarrow e \nu_{\tau} \overline{\nu_{e}}\right) \simeq 0.18$ and $\BR\left(\tau \rightarrow \mu \nu_{\tau} \overline{\nu_{\mu}}\right) \simeq 0.17$~\cite{ParticleDataGroup:2020ssz}. The mixing matrix $\mathbf{K}$ and neutrino masses $m_{i}$ are given by Eqs.~\eqref{eq:K} and~\eqref{eq:Unudef}, respectively, while $\mathbf{Y}_f$ and $M_f$ have been defined in Eq.~\eqref{eq:LYuk}. The \emph{dark} charged-scalar mass $m_{\zeta^{+}}$ is given by Eq.~\eqref{eq:mcharged} and the loop functions $G_W(x)$~\cite{Cheng:1980tp,Lim:1981kv,Langacker:1988up,Ilakovac:1994kj} and $G_{\zeta^{+}}(x)$~\cite{Toma:2013zsa,Vicente:2014wga} are
\begin{align}
G_W(x) = \frac{x (1-5x-2x^2)}{4 (1-x)^3} - \frac{3 x^3}{2 (1-x)^4} \ln x \; ,  \; G_{\zeta^{+}}(x) = \frac{1 - 5 x - 2 x^2}{6 (1-x)^3} - \frac{x^2 \ln x}{(1-x)^4} \; .
\end{align}

Our numerical results are presented in Fig.~\ref{fig:cLFV}. In the upper plot, we show $|G^{\mu e}_{\zeta^+}|$ versus $|G^{\mu e}_{W}|$ [see Eq.~\eqref{eq:cLFV}]. Along the brown-dashed contour $|G^{\mu e}_{\zeta^+}| = |G^{\mu e}_{W}|$, above which the \emph{dark} (seesaw) contribution dominates, i.e. $|G^{\mu e}_{\zeta^+}| > |G^{\mu e}_{W}|$ ($|G^{\mu e}_{\zeta^+}| < |G^{\mu e}_{W}|$). The red points are excluded by the current MEG bound BR$(\mu \rightarrow e \gamma) < 4.2 \times 10^{-13}$~\cite{MEG:2016leq}, while the green ones are within the sensitivity reach of MEG II~\cite{MEGII:2018kmf}. Cancellations between $G^{\mu e}_{\zeta^+}$ and $G^{\mu e}_{W}$ along the brown-dashed line allow for small BRs with large $|G^{\mu e}_{\zeta^+}|$ and $|G^{\mu e}_{W}|$. In the lower plots, we present $\BR(\mu \rightarrow e \gamma)$ versus $\BR(\tau \rightarrow e \gamma)$ (left) and $\BR(\tau \rightarrow \mu \gamma)$ (right). It is apparent from our results that once the current MEG constraint is applied (horizontal red-solid line), the constraints on the $\tau$ radiative processes stemming from BaBar~\cite{BaBar:2009hkt} (current - vertical red-solid line) are automatically satisfied, leading to decay rates beyond the sensitivity reach of Belle II~\cite{Belle-II:2018jsg} (future - vertical green-dashed line).

\section{Concluding remarks}
\label{sec:concl}

We have presented a minimal model where neutrino masses stem from the linear seesaw mechanism radiatively ``seeded'' by a \emph{dark} sector. Namely, the required small LNV parameter arises at one-loop mediated by viable DM candidates. We have considered the simplest setup containing only one $\nu_R-S_R$ heavy neutrino pair, as well as a single \emph{dark} vector-like fermion $f_{L,R}$. This is sufficient to explain the two neutrino mass splittings, predicting a massless light neutrino. Although the model may accommodate either fermionic or scalar DM, we focused on the latter possibility with DM being a mixed state of SU$(2)_{\text{L}}$ singlet and doublet scalars. Due to this feature, DM with the right relic abundance can be obtained for a wide mass range of $\sim [45~{\rm GeV},2~{\rm TeV}]$. This scenario can also be directly probed at future DD experiments such as LZ, XENONnT and DARWIN. Being a low-scale seesaw scheme, the \emph{dark}-seeded linear seesaw leads to sizeable cLFV signatures making the seesaw paradigm testable at future cLFV facilities like MEG II and Belle II, with distinct contributions from the \emph{dark} sector.

Overall, our setup is the simplest framework in which a \emph{dark}-sector induces the small LNV parameter generating non-zero neutrino masses via the linear seesaw mechanism, without contributing to any other mass term which could generate light neutrino masses through another mechanism. This feature is due to the Dirac character of $f_{L,R}$. Other approaches have been followed in the literature, for instance in the context of gauged $B-L$ symmetry~\cite{Wang:2015saa,Das:2017ski} or in the $331$ model~\cite{CarcamoHernandez:2021tlv}, inevitably requiring more involved particle content and symmetries such as multiple copies of sterile singlet fermions, vector-like charged leptons for anomaly cancellation or non-minimal scalar sectors. These examples lead to distinct phenomenological signatures in regards to DM and cLFV, involving annihilations of dark matter candidates to new vector bosons. Lastly, we wish to point out that the philosophy of connecting DM with LNV has also been applied in the context of other low-scale seesaw schemes such as the inverse seesaw~\cite{Ma:2009gu,Bazzocchi:2009kc,Baldes:2013eva,CarcamoHernandez:2018hst,Mandal:2019oth}.\\

\noindent {\bf NOTE:} While this Letter was at its final-reading stage, we came to know that J.~Valle and collaborators were preparing a paper about an alternative realization of the same idea presented here.

\begin{acknowledgments}
We thank J.~Valle and R.~Srivastava for reading the manuscript. This work is supported by Fundação para a Ciência e a Tecnologia (FCT, Portugal) through the projects CFTP-FCT Unit UIDB/00777/2020 and UIDP/00777/2020, CERN/FIS-PAR/0019/2021, which are partially funded through POCTI (FEDER), COMPETE, QREN and EU. The work of A.B. and H.C. is supported by the PhD FCT grants UI/BD/154391/2023 and 2021.06340.BD, respectively.
\end{acknowledgments}

\appendix

\section{Scalar sector and \emph{dark}-loop functions}
\label{sec:scalar}

As shown in Table~\ref{tab:model}, the scalar content of our model contains, besides the usual Higgs doublet $\Phi$, a doublet $\eta$, which we define as
\begin{align}
\Phi&=\begin{pmatrix}
\phi^{+} \\
\phi^0
\end{pmatrix}= \frac{1}{\sqrt{2}}  \begin{pmatrix}
 \sqrt{2} \phi^{+} \\
 v + \phi_{\text{R}} + i \phi_{\text{I}}
\end{pmatrix} \; ; \; \eta =\begin{pmatrix}
\eta^{+} \\
\eta^0
\end{pmatrix}= \frac{1}{\sqrt{2}}  \begin{pmatrix}
 \sqrt{2} \eta^{+} \\
\eta_{\text{R}} + i \eta_{\text{I}}
\end{pmatrix} \; ;
\label{eq:scalars}
\end{align}
as well as, a real scalar singlet $\chi$. The Higgs doublet acquires a non-zero VEV where $\left< \phi^0 \right> = v/\sqrt{2} \simeq 174$ GeV, with the doublet $\eta$ and singlet $\chi$ being inert, i.e. $\left< \eta^0 \right> = \left< \chi \right> = 0$. The full scalar potential $V(\Phi,\eta,\chi)$ allowed by the symmetries of our model is,
    \begin{align}
    V &=  m_{\Phi}^2 \left(\Phi^\dagger \Phi\right) + m_{\eta}^2 \left(\eta^\dagger \eta\right) + m_\chi^2 \chi^2 + \frac{\lambda_{\Phi}}{2} \left(\Phi^\dagger \Phi\right)^2  + \lambda_{3} \left(\Phi^\dagger \Phi\right) \left(\eta^\dagger \eta\right) + \lambda_{4} \left(\Phi^\dagger \eta \right) \left(\eta^\dagger \Phi\right) \nonumber \\
    & + \frac{\lambda_{\eta}}{2} \left(\eta^\dagger \eta\right)^2 + \frac{\lambda_{\chi}}{2} \chi^4 + \lambda_{\Phi \chi} \left(\Phi^\dagger \Phi\right) \chi^2 +  \lambda_{\eta \chi} \left(\eta^\dagger \eta\right) \chi^2 + \left[\kappa \left(\eta^\dagger \Phi \right) \chi + \text{H.c.} \right]\; ,
    \label{eq:Vpotentialfull}
    \end{align}
    where $\kappa$ is a soft LNV parameter which can be made real. In order for the potential to be bounded from below, the quartic parameters must obey the following conditions:
    \begin{align}
    &\lambda_{\Phi},\lambda_{\eta},\lambda_{\chi} > 0\,, \,\lambda_{3} + \sqrt{\lambda_\Phi \lambda_\eta} > 0\;,\;\lambda_{3} + \lambda_{4} + \sqrt{\lambda_\Phi \lambda_\eta} > 0\;,\;\lambda_{\Phi \chi} + \sqrt{\lambda_\Phi \lambda_{\chi}} > 0\;,\;\lambda_{\eta \chi} + \sqrt{\lambda_\eta \lambda_{\chi}} > 0 \; , \label{eq:bounded} \\
    &\lambda_{3}\lambda_{\chi} - \lambda_{\Phi \chi} \lambda_{\eta \chi} + \sqrt{(\lambda_\phi \lambda_{\chi} - \lambda_{\Phi \chi}^2) (\lambda_\eta \lambda_{\chi} - \lambda_{\eta \chi}^2)} > 0\;, \; (\lambda_{3} + \lambda_4)\lambda_{\chi} - \lambda_{\Phi \chi} \lambda_{\eta \chi} + \sqrt{(\lambda_\phi \lambda_{\chi} - \lambda_{\Phi \chi}^2) (\lambda_\eta \lambda_{\chi} - \lambda_{\eta \chi}^2) } > 0\;. \nonumber
    \end{align}

The \textit{dark}-scalar sector contains a charged $\zeta^\pm$ and three neutral $\zeta_k$ ($k=1, 2, 3$) states. The former is relevant for cLFV (see section~\ref{sec:cLFV}) while the latter generate neutrino masses~(see section~\ref{sec:model}). The \textit{dark}-charged scalar mass is
\begin{equation}
m_{\zeta^+}^2 = m_\eta^2 + \frac{1}{2} v^2 \lambda_3 \; , \label{eq:mcharged}
\end{equation}
and the mass matrix for the \textit{dark}-neutral scalar states in the $\left(\eta_\text{R}, \chi, \eta_{\text{I}} \right)$ basis reads,
\begin{align}
    \mathcal{M}^2_{\eta \chi}= \begin{pmatrix}
   m_\eta^2 + \frac{v^2}{2} \left(\lambda_3+\lambda_4\right) 
    & v \kappa & 0 \\
    \cdot&2 m_\chi^2 + \lambda_{\Phi\chi} v^2 &  0\\
    \cdot& \cdot &   m_\eta^2 + \frac{v^2}{2} \left(\lambda_3+\lambda_4\right)
    \end{pmatrix} \; ,
    \label{eq:darkscalarmass}
\end{align}
where '$\cdot$' reflects the symmetric nature of the matrix. The neutral components of the $\eta$ and $\chi$ are related to the mass-eigenstates~$\zeta_{i}$ through the $3 \times 3$ orthogonal matrix $\mathbf{V}$:
\begin{equation}
(\eta_{\text{R}},
\chi,
\eta_{\text{I}})^T
 = \mathbf{V} 
(\zeta_1, \zeta_2, \zeta_3)^T \; , \; \sqrt{2} \eta^0 = \sum_{k=1}^{3} \left(\mathbf{V}_{1 k} + i \mathbf{V}_{3 k} \right) \zeta_k \; , \; \chi = \sum_{k=1}^{3} \mathbf{V}_{2 k} \zeta_k \; ,
\label{eq:mixneutralscotoDM}
\end{equation}
The $\zeta_{i}$ masses are,
\begin{align}
m_{\zeta_1}^2 = m_\eta^2 + \frac{v^2}{2} \left(\lambda_3+\lambda_4\right) \; , \; m_{\zeta_2}^2 = m_{\zeta_0}^2 - \frac{1}{4} \sqrt{\Lambda} \; , \; m_{\zeta_3}^2 = m_{\zeta_0}^2 + \frac{1}{4} \sqrt{\Lambda} \; ,
\end{align}
with,
\begin{align}
m_{\zeta_0}^2 = \frac{1}{2} \left(m_\eta^2 +2 m_\chi^2 \right) + \frac{v^2}{4} \left(\lambda_3+\lambda_4+ 2 \lambda_{\Phi \chi} \right) \; , \;
\Lambda = \left[2 \left(m_\eta^2 - 2 m_\chi^2 \right) + v^2 \left(\lambda_3+\lambda_4-2\lambda_{\Phi\chi} \right) \right]^2 + 16 v^2 \kappa^2 \; .
\end{align}
The \emph{dark} one-loop functions relevant for neutrino mass generation are given by,
\begin{align}
\mathcal{F}_{S} \left(M_f , m_{\zeta_k}\right) = & \; \frac{1}{16 \sqrt{2} \pi^2} \sum_{k=1}^{3} \left(\mathbf{V}_{1 k} - i \mathbf{V}_{3 k} \right) \mathbf{V}_{2 k} \frac{m_{\zeta_k}^2}{M_f^2 - m_{\zeta_k}^2} \ln \left( \frac{M_f^2}{m_{\zeta_k}^2} \right) \; ,
\nonumber \\
\mathcal{F}_{B} \left(M_f , m_{\zeta_k}\right) = & \; \frac{1}{16 \pi^2} \sum_{k=1}^{3} \mathbf{V}_{2 k}^2 \frac{m_{\zeta_k}^2}{M_f^2 - m_{\zeta_k}^2} \ln \left( \frac{M_f^2}{m_{\zeta_k}^2} \right) \; .
\label{eq:Floop}
\end{align}

\section{Generalised Casas-Ibarra parameterisation}
\label{sec:CI}

A convenient way to obtain, neutrino data compatible, Yukawa coupling vectors $\mathbf{Y}_{D,f}$ of Eq.~\eqref{eq:LYuk}, is via the  Casas-Ibarra parameterisation~\cite{Casas:2001sr}.  We start by writing $\mathbf{M}_\nu$ of Eq.~\eqref{eq:Mnu} as follows,
\begin{equation}
  \mathbf{M}_{\nu} = - \frac{v^2}{2} \mathbf{Y} \; \mathbf{M}^{-1} \mathbf{Y}^T \; ,\; \mathbf{Y} = \left(\mathbf{Y}_{D} \; \mathbf{Y}_{f}\right) \; , \; \mathbf{M} = \begin{pmatrix}
      0 & |\tilde{M}| e^{i \phi}\\
      |\tilde{M}| e^{i \phi} & 0
  \end{pmatrix} \; , 
\end{equation}
where we define the $3\times 2$ Yukawa matrix $\mathbf{Y}$, the $2\times 2$ mass matrix $\mathbf{M}$ and
\begin{align}
    \tilde{M} = v \frac{M_B+\mathcal{F}_{B} M_f \; Y_{R} Y_S}{M_f} \frac{1}{\sqrt{2} Y_S \mathcal{F}_S} \; , \; \phi = \arg \tilde{M} \; ,
\end{align}
with the masses $M_B,M_f$ and Yukawa parameters $Y_R,Y_{S}$ and \emph{dark}-loop functions $\mathcal{F}$ given in Eqs.~\eqref{eq:LYuk} and~\eqref{eq:Floop}, respectively. The matrix $\mathbf{M}$ can be diagonalised as
\begin{align}
\U_M^{\dagger}\, \M\, \U_M^\ast &= \mathbf{d}_{M} = \text{diag}\left(|\tilde{M}| , |\tilde{M}| \right) \; , \;
\U_M = \frac{e^{i \phi}}{\sqrt{2}} \begin{pmatrix}
    i & 1 \\
    -i & 1
\end{pmatrix} \; . 
\label{eq:Mheavy}
\end{align}
From the above we find the following parameterisation for $\mathbf{Y}$:
\begin{equation} \label{eq:CI}
    \mathbf{Y} = i \frac{\sqrt{2}}{v} \mathbf{U}_{\nu}^\ast\, \mathbf{d}_\nu^{1/2}\, \mathbf{R}\, \mathbf{d}_M^{1/2} \mathbf{U}_M^T \; , \; \mathbf{R}_\text{NO} = \begin{pmatrix}
        0 & 0 \\
        \cos (z) & - \sin (z) \\
        \pm \sin (z) & \pm \cos (z)
    \end{pmatrix} \; , \; 
   \mathbf{R}_\text{IO} = \begin{pmatrix}
        \cos (z) & - \sin (z) \\
        \pm \sin (z) & \pm \cos (z) \\
         0 & 0
    \end{pmatrix} \,,
\end{equation}
where $\mathbf{d}_\nu$, $\mathbf{U}_{\nu}$, $\mathbf{d}_M$ and $\mathbf{U}_M$ are given by Eqs.~\eqref{eq:Unudef} and~\eqref{eq:Mheavy}, respectively. Furthermore, $\mathbf{R}$ is an orthogonal $3\times 2$ matrix
with $z$ being a complex parameter.


\begin{thebibliography}{10}

\bibitem{Kajita:2016cak}
T.~Kajita, ``{Nobel Lecture: Discovery of atmospheric neutrino oscillations},''
  \href{http://dx.doi.org/10.1103/RevModPhys.88.030501}{{\em Rev. Mod. Phys.}
  {\bfseries 88} no.~3, (2016) 030501}.

\bibitem{McDonald:2016ixn}
A.~B. McDonald, ``{Nobel Lecture: The Sudbury Neutrino Observatory: Observation
  of flavor change for solar neutrinos},''
  \href{http://dx.doi.org/10.1103/RevModPhys.88.030502}{{\em Rev. Mod. Phys.}
  {\bfseries 88} no.~3, (2016) 030502}.

\bibitem{Bertone:2004pz}
G.~Bertone, D.~Hooper, and J.~Silk, ``{Particle dark matter: Evidence,
  candidates and constraints},''
  \href{http://dx.doi.org/10.1016/j.physrep.2004.08.031}{{\em Phys. Rept.}
  {\bfseries 405} (2005) 279--390},
  \href{http://arxiv.org/abs/hep-ph/0404175}{{\ttfamily arXiv:hep-ph/0404175}}.

\bibitem{Planck:2018vyg}
{\bfseries Planck} Collaboration, N.~Aghanim {\em et~al.}, ``{Planck 2018
  results. VI. Cosmological parameters},''
  \href{http://dx.doi.org/10.1051/0004-6361/201833910}{{\em Astron. Astrophys.}
  {\bfseries 641} (2020) A6}, \href{http://arxiv.org/abs/1807.06209}{{\ttfamily
  arXiv:1807.06209 [astro-ph.CO]}}. [Erratum: Astron.Astrophys. 652, C4
  (2021)].

\bibitem{Minkowski:1977sc}
P.~Minkowski, ``{$\mu \to e\gamma$ at a Rate of One Out of $10^{9}$ Muon
  Decays?},'' \href{http://dx.doi.org/10.1016/0370-2693(77)90435-X}{{\em Phys.
  Lett. B} {\bfseries 67} (1977) 421--428}.

\bibitem{Gell-Mann:1979vob}
M.~Gell-Mann, P.~Ramond, and R.~Slansky, ``{Complex Spinors and Unified
  Theories},'' {\em Conf. Proc. C} {\bfseries 790927} (1979) 315--321,
  \href{http://arxiv.org/abs/1306.4669}{{\ttfamily arXiv:1306.4669 [hep-th]}}.

\bibitem{Yanagida:1979as}
T.~Yanagida, ``{Horizontal gauge symmetry and masses of neutrinos},'' {\em
  Conf. Proc. C} {\bfseries 7902131} (1979) 95--99.

\bibitem{Schechter:1980gr}
J.~Schechter and J.~W.~F. Valle, ``{Neutrino Masses in SU(2) x U(1)
  Theories},'' \href{http://dx.doi.org/10.1103/PhysRevD.22.2227}{{\em Phys.
  Rev. D} {\bfseries 22} (1980) 2227}.

\bibitem{Glashow:1979nm}
S.~L. Glashow, ``{The Future of Elementary Particle Physics},''
  \href{http://dx.doi.org/10.1007/978-1-4684-7197-7_15}{{\em NATO Sci. Ser. B}
  {\bfseries 61} (1980) 687}.

\bibitem{Mohapatra:1979ia}
R.~N. Mohapatra and G.~Senjanovic, ``{Neutrino Mass and Spontaneous Parity
  Nonconservation},'' \href{http://dx.doi.org/10.1103/PhysRevLett.44.912}{{\em
  Phys. Rev. Lett.} {\bfseries 44} (1980) 912}.

\bibitem{Zee:1980ai}
A.~Zee, ``{A Theory of Lepton Number Violation, Neutrino Majorana Mass, and
  Oscillation},'' \href{http://dx.doi.org/10.1016/0370-2693(80)90349-4}{{\em
  Phys. Lett. B} {\bfseries 93} (1980) 389}. [Erratum: Phys.Lett.B 95, 461
  (1980)].

\bibitem{Ma:1998dn}
E.~Ma, ``{Pathways to naturally small neutrino masses},''
  \href{http://dx.doi.org/10.1103/PhysRevLett.81.1171}{{\em Phys. Rev. Lett.}
  {\bfseries 81} (1998) 1171--1174},
  \href{http://arxiv.org/abs/hep-ph/9805219}{{\ttfamily arXiv:hep-ph/9805219}}.

\bibitem{Mohapatra:1986bd}
R.~N. Mohapatra and J.~W.~F. Valle, ``{Neutrino Mass and Baryon Number
  Nonconservation in Superstring Models},''
  \href{http://dx.doi.org/10.1103/PhysRevD.34.1642}{{\em Phys. Rev. D}
  {\bfseries 34} (1986) 1642}.

\bibitem{Akhmedov:1995ip}
E.~K. Akhmedov, M.~Lindner, E.~Schnapka, and J.~W.~F. Valle, ``{Left-right
  symmetry breaking in NJL approach},''
  \href{http://dx.doi.org/10.1016/0370-2693(95)01504-3}{{\em Phys. Lett. B}
  {\bfseries 368} (1996) 270--280},
  \href{http://arxiv.org/abs/hep-ph/9507275}{{\ttfamily arXiv:hep-ph/9507275}}.

\bibitem{Akhmedov:1995vm}
E.~K. Akhmedov, M.~Lindner, E.~Schnapka, and J.~W.~F. Valle, ``{Dynamical
  left-right symmetry breaking},''
  \href{http://dx.doi.org/10.1103/PhysRevD.53.2752}{{\em Phys. Rev. D}
  {\bfseries 53} (1996) 2752--2780},
  \href{http://arxiv.org/abs/hep-ph/9509255}{{\ttfamily arXiv:hep-ph/9509255}}.

\bibitem{Malinsky:2005bi}
M.~Malinsky, J.~C. Romao, and J.~W.~F. Valle, ``{Novel supersymmetric SO(10)
  seesaw mechanism},''
  \href{http://dx.doi.org/10.1103/PhysRevLett.95.161801}{{\em Phys. Rev. Lett.}
  {\bfseries 95} (2005) 161801},
  \href{http://arxiv.org/abs/hep-ph/0506296}{{\ttfamily arXiv:hep-ph/0506296}}.

\bibitem{Alonso:2012ji}
R.~Alonso, M.~Dhen, M.~B. Gavela, and T.~Hambye, ``{Muon conversion to electron
  in nuclei in type-I seesaw models},''
  \href{http://dx.doi.org/10.1007/JHEP01(2013)118}{{\em JHEP} {\bfseries 01}
  (2013) 118}, \href{http://arxiv.org/abs/1209.2679}{{\ttfamily arXiv:1209.2679
  [hep-ph]}}.

\bibitem{Abada:2015oba}
A.~Abada, V.~De~Romeri, and A.~M. Teixeira, ``{Impact of sterile neutrinos on
  nuclear-assisted cLFV processes},''
  \href{http://dx.doi.org/10.1007/JHEP02(2016)083}{{\em JHEP} {\bfseries 02}
  (2016) 083}, \href{http://arxiv.org/abs/1510.06657}{{\ttfamily
  arXiv:1510.06657 [hep-ph]}}.

\bibitem{Abada:2018nio}
A.~Abada and A.~M. Teixeira, ``{Heavy neutral leptons and high-intensity
  observables},'' \href{http://dx.doi.org/10.3389/fphy.2018.00142}{{\em Front.
  in Phys.} {\bfseries 6} (2018) 142},
  \href{http://arxiv.org/abs/1812.08062}{{\ttfamily arXiv:1812.08062
  [hep-ph]}}.

\bibitem{Hernandez-Tome:2019lkb}
G.~Hern\'andez-Tom\'e, J.~I. Illana, M.~Masip, G.~L\'opez~Castro, and P.~Roig,
  ``{Effects of heavy Majorana neutrinos on lepton flavor violating
  processes},'' \href{http://dx.doi.org/10.1103/PhysRevD.101.075020}{{\em Phys.
  Rev. D} {\bfseries 101} no.~7, (2020) 075020},
  \href{http://arxiv.org/abs/1912.13327}{{\ttfamily arXiv:1912.13327
  [hep-ph]}}.

\bibitem{Camara:2020efq}
H.~B. Camara, R.~G. Felipe, and F.~R. Joaquim, ``{Minimal inverse-seesaw
  mechanism with Abelian flavour symmetries},''
  \href{http://dx.doi.org/10.1007/JHEP05(2021)021}{{\em JHEP} {\bfseries 05}
  (2021) 021}, \href{http://arxiv.org/abs/2012.04557}{{\ttfamily
  arXiv:2012.04557 [hep-ph]}}.

\bibitem{Ma:2006km}
E.~Ma, ``{Verifiable radiative seesaw mechanism of neutrino mass and dark
  matter},'' \href{http://dx.doi.org/10.1103/PhysRevD.73.077301}{{\em Phys.
  Rev. D} {\bfseries 73} (2006) 077301},
  \href{http://arxiv.org/abs/hep-ph/0601225}{{\ttfamily arXiv:hep-ph/0601225}}.

\bibitem{Cai:2017jrq}
Y.~Cai, J.~Herrero-Garc\'\i{}a, M.~A. Schmidt, A.~Vicente, and R.~R. Volkas,
  ``{From the trees to the forest: a review of radiative neutrino mass
  models},'' \href{http://dx.doi.org/10.3389/fphy.2017.00063}{{\em Front. in
  Phys.} {\bfseries 5} (2017) 63},
  \href{http://arxiv.org/abs/1706.08524}{{\ttfamily arXiv:1706.08524
  [hep-ph]}}.

\bibitem{tHooft:1979rat}
G.~'t~Hooft, ``{Naturalness, chiral symmetry, and spontaneous chiral symmetry
  breaking},'' \href{http://dx.doi.org/10.1007/978-1-4684-7571-5_9}{{\em NATO
  Sci. Ser. B} {\bfseries 59} (1980) 135--157}.

\bibitem{Schechter:1981cv}
J.~Schechter and J.~W.~F. Valle, ``{Neutrino Decay and Spontaneous Violation of
  Lepton Number},'' \href{http://dx.doi.org/10.1103/PhysRevD.25.774}{{\em Phys.
  Rev. D} {\bfseries 25} (1982) 774}.

\bibitem{ParticleDataGroup:2020ssz}
{\bfseries Particle Data Group} Collaboration, P.~A. Zyla {\em et~al.},
  ``{Review of Particle Physics},''
  \href{http://dx.doi.org/10.1093/ptep/ptaa104}{{\em PTEP} {\bfseries 2020}
  no.~8, (2020) 083C01}.

\bibitem{Schechter:1981bd}
J.~Schechter and J.~W.~F. Valle, ``{Neutrinoless Double beta Decay in SU(2) x
  U(1) Theories},'' \href{http://dx.doi.org/10.1103/PhysRevD.25.2951}{{\em
  Phys. Rev. D} {\bfseries 25} (1982) 2951}.

\bibitem{Barreiros:2018bju}
D.~M. Barreiros, R.~G. Felipe, and F.~R. Joaquim, ``{Combining texture zeros
  with a remnant CP symmetry in the minimal type-I seesaw},''
  \href{http://dx.doi.org/10.1007/JHEP01(2019)223}{{\em JHEP} {\bfseries 01}
  (2019) 223}, \href{http://arxiv.org/abs/1810.05454}{{\ttfamily
  arXiv:1810.05454 [hep-ph]}}.

\bibitem{SNO:2015wyx}
{\bfseries SNO+} Collaboration, S.~Andringa {\em et~al.}, ``{Current Status and
  Future Prospects of the SNO+ Experiment},''
  \href{http://dx.doi.org/10.1155/2016/6194250}{{\em Adv. High Energy Phys.}
  {\bfseries 2016} (2016) 6194250},
  \href{http://arxiv.org/abs/1508.05759}{{\ttfamily arXiv:1508.05759
  [physics.ins-det]}}.

\bibitem{LEGEND:2017cdu}
{\bfseries LEGEND} Collaboration, N.~Abgrall {\em et~al.}, ``{The Large
  Enriched Germanium Experiment for Neutrinoless Double Beta Decay (LEGEND)},''
  \href{http://dx.doi.org/10.1063/1.5007652}{{\em AIP Conf. Proc.} {\bfseries
  1894} no.~1, (2017) 020027},
  \href{http://arxiv.org/abs/1709.01980}{{\ttfamily arXiv:1709.01980
  [physics.ins-det]}}.

\bibitem{nEXO:2017nam}
{\bfseries nEXO} Collaboration, J.~B. Albert {\em et~al.}, ``{Sensitivity and
  Discovery Potential of nEXO to Neutrinoless Double Beta Decay},''
  \href{http://dx.doi.org/10.1103/PhysRevC.97.065503}{{\em Phys. Rev. C}
  {\bfseries 97} no.~6, (2018) 065503},
  \href{http://arxiv.org/abs/1710.05075}{{\ttfamily arXiv:1710.05075
  [nucl-ex]}}.

\bibitem{deSalas:2020pgw}
P.~F. de~Salas, D.~V. Forero, S.~Gariazzo, P.~Mart\'\i{}nez-Mirav\'e, O.~Mena,
  C.~A. Ternes, M.~T\'ortola, and J.~W.~F. Valle, ``{2020 global reassessment
  of the neutrino oscillation picture},''
  \href{http://dx.doi.org/10.1007/JHEP02(2021)071}{{\em JHEP} {\bfseries 02}
  (2021) 071}, \href{http://arxiv.org/abs/2006.11237}{{\ttfamily
  arXiv:2006.11237 [hep-ph]}}.

\bibitem{Staub:2013tta}
F.~Staub, ``{SARAH 4 : A tool for (not only SUSY) model builders},''
  \href{http://dx.doi.org/10.1016/j.cpc.2014.02.018}{{\em Comput.Phys.Commun.}
  {\bfseries 185} (2014) 1773--1790},
  \href{http://arxiv.org/abs/1309.7223}{{\ttfamily arXiv:1309.7223 [hep-ph]}}.

\bibitem{Porod:2003um}
W.~Porod, ``{SPheno, a program for calculating supersymmetric spectra, SUSY
  particle decays and SUSY particle production at e+ e- colliders},''
  \href{http://dx.doi.org/10.1016/S0010-4655(03)00222-4}{{\em
  Comput.Phys.Commun.} {\bfseries 153} (2003) 275--315},
  \href{http://arxiv.org/abs/hep-ph/0301101}{{\ttfamily arXiv:hep-ph/0301101
  [hep-ph]}}.

\bibitem{Belanger:2014vza}
G.~B{\'e}langer, F.~Boudjema, A.~Pukhov, and A.~Semenov, ``{micrOMEGAs4.1: two
  dark matter candidates},''
  \href{http://dx.doi.org/10.1016/j.cpc.2015.03.003}{{\em Comput.Phys.Commun.}
  {\bfseries 192} (2015) 322--329},
  \href{http://arxiv.org/abs/1407.6129}{{\ttfamily arXiv:1407.6129 [hep-ph]}}.

\bibitem{Cao:2007rm}
Q.-H. Cao, E.~Ma, and G.~Rajasekaran, ``{Observing the Dark Scalar Doublet and
  its Impact on the Standard-Model Higgs Boson at Colliders},''
  \href{http://dx.doi.org/10.1103/PhysRevD.76.095011}{{\em Phys. Rev. D}
  {\bfseries 76} (2007) 095011},
  \href{http://arxiv.org/abs/0708.2939}{{\ttfamily arXiv:0708.2939 [hep-ph]}}.

\bibitem{Gustafsson:2007pc}
M.~Gustafsson, E.~Lundstrom, L.~Bergstrom, and J.~Edsjo, ``{Significant Gamma
  Lines from Inert Higgs Dark Matter},''
  \href{http://dx.doi.org/10.1103/PhysRevLett.99.041301}{{\em Phys. Rev. Lett.}
  {\bfseries 99} (2007) 041301},
  \href{http://arxiv.org/abs/astro-ph/0703512}{{\ttfamily
  arXiv:astro-ph/0703512}}.

\bibitem{Pierce:2007ut}
A.~Pierce and J.~Thaler, ``{Natural Dark Matter from an Unnatural Higgs Boson
  and New Colored Particles at the TeV Scale},''
  \href{http://dx.doi.org/10.1088/1126-6708/2007/08/026}{{\em JHEP} {\bfseries
  08} (2007) 026}, \href{http://arxiv.org/abs/hep-ph/0703056}{{\ttfamily
  arXiv:hep-ph/0703056}}.

\bibitem{ATLAS:2016neq}
{\bfseries ATLAS, CMS} Collaboration, G.~Aad {\em et~al.}, ``{Measurements of
  the Higgs boson production and decay rates and constraints on its couplings
  from a combined ATLAS and CMS analysis of the LHC pp collision data at $
  \sqrt{s}=7 $ and 8 TeV},''
  \href{http://dx.doi.org/10.1007/JHEP08(2016)045}{{\em JHEP} {\bfseries 08}
  (2016) 045}, \href{http://arxiv.org/abs/1606.02266}{{\ttfamily
  arXiv:1606.02266 [hep-ex]}}.

\bibitem{ATLAS:2019nkf}
{\bfseries ATLAS} Collaboration, G.~Aad {\em et~al.}, ``{Combined measurements
  of Higgs boson production and decay using up to $80$ fb$^{-1}$ of
  proton-proton collision data at $\sqrt{s}=$ 13 TeV collected with the ATLAS
  experiment},'' \href{http://dx.doi.org/10.1103/PhysRevD.101.012002}{{\em
  Phys. Rev. D} {\bfseries 101} no.~1, (2020) 012002},
  \href{http://arxiv.org/abs/1909.02845}{{\ttfamily arXiv:1909.02845
  [hep-ex]}}.

\bibitem{CMS:2018yfx}
{\bfseries CMS} Collaboration, A.~M. Sirunyan {\em et~al.}, ``{Search for
  invisible decays of a Higgs boson produced through vector boson fusion in
  proton-proton collisions at $\sqrt{s} =$ 13 TeV},''
  \href{http://dx.doi.org/10.1016/j.physletb.2019.04.025}{{\em Phys. Lett. B}
  {\bfseries 793} (2019) 520--551},
  \href{http://arxiv.org/abs/1809.05937}{{\ttfamily arXiv:1809.05937
  [hep-ex]}}.

\bibitem{ATLAS:2018hxb}
{\bfseries ATLAS} Collaboration, M.~Aaboud {\em et~al.}, ``{Measurements of
  Higgs boson properties in the diphoton decay channel with 36 fb$^{-1}$ of
  $pp$ collision data at $\sqrt{s} = 13$ TeV with the ATLAS detector},''
  \href{http://dx.doi.org/10.1103/PhysRevD.98.052005}{{\em Phys. Rev. D}
  {\bfseries 98} (2018) 052005},
  \href{http://arxiv.org/abs/1802.04146}{{\ttfamily arXiv:1802.04146
  [hep-ex]}}.

\bibitem{MEG:2016leq}
{\bfseries MEG} Collaboration, A.~M. Baldini {\em et~al.}, ``{Search for the
  lepton flavour violating decay $\mu ^+ \rightarrow \mathrm {e}^+ \gamma $
  with the full dataset of the MEG experiment},''
  \href{http://dx.doi.org/10.1140/epjc/s10052-016-4271-x}{{\em Eur. Phys. J. C}
  {\bfseries 76} no.~8, (2016) 434},
  \href{http://arxiv.org/abs/1605.05081}{{\ttfamily arXiv:1605.05081
  [hep-ex]}}.

\bibitem{LZ:2022ufs}
{\bfseries LZ} Collaboration, J.~Aalbers {\em et~al.}, ``{First Dark Matter
  Search Results from the LUX-ZEPLIN (LZ) Experiment},''
  \href{http://arxiv.org/abs/2207.03764}{{\ttfamily arXiv:2207.03764
  [hep-ex]}}.

\bibitem{LZ:2018qzl}
{\bfseries LUX-ZEPLIN} Collaboration, D.~Akerib {\em et~al.}, ``{Projected WIMP
  sensitivity of the LUX-ZEPLIN dark matter experiment},''
  \href{http://dx.doi.org/10.1103/PhysRevD.101.052002}{{\em Phys.Rev.D}
  {\bfseries 101} (2020) 052002},
  \href{http://arxiv.org/abs/1802.06039}{{\ttfamily arXiv:1802.06039
  [astro-ph.IM]}}.

\bibitem{XENON:2020kmp}
{\bfseries XENON} Collaboration, E.~Aprile {\em et~al.}, ``{Projected WIMP
  sensitivity of the XENONnT dark matter experiment},''
  \href{http://dx.doi.org/10.1088/1475-7516/2020/11/031}{{\em JCAP} {\bfseries
  11} (2020) 031}, \href{http://arxiv.org/abs/2007.08796}{{\ttfamily
  arXiv:2007.08796 [physics.ins-det]}}.

\bibitem{DARWIN:2016hyl}
{\bfseries DARWIN} Collaboration, J.~Aalbers {\em et~al.}, ``{DARWIN: towards
  the ultimate dark matter detector},''
  \href{http://dx.doi.org/10.1088/1475-7516/2016/11/017}{{\em JCAP} {\bfseries
  11} (2016) 017}, \href{http://arxiv.org/abs/1606.07001}{{\ttfamily
  arXiv:1606.07001 [astro-ph.IM]}}.

\bibitem{Billard:2013qya}
J.~Billard, L.~Strigari, and E.~Figueroa-Feliciano, ``{Implication of neutrino
  backgrounds on the reach of next generation dark matter direct detection
  experiments},'' \href{http://dx.doi.org/10.1103/PhysRevD.89.023524}{{\em
  Phys.Rev.D} {\bfseries 89} (2014) 023524},
  \href{http://arxiv.org/abs/1307.5458}{{\ttfamily arXiv:1307.5458 [hep-ph]}}.

\bibitem{McDonald:1993ex}
J.~McDonald, ``{Gauge singlet scalars as cold dark matter},''
  \href{http://dx.doi.org/10.1103/PhysRevD.50.3637}{{\em Phys. Rev. D}
  {\bfseries 50} (1994) 3637--3649},
  \href{http://arxiv.org/abs/hep-ph/0702143}{{\ttfamily arXiv:hep-ph/0702143}}.

\bibitem{Guo:2010hq}
W.-L. Guo and Y.-L. Wu, ``{The Real singlet scalar dark matter model},''
  \href{http://dx.doi.org/10.1007/JHEP10(2010)083}{{\em JHEP} {\bfseries 10}
  (2010) 083}, \href{http://arxiv.org/abs/1006.2518}{{\ttfamily arXiv:1006.2518
  [hep-ph]}}.

\bibitem{Cline:2013gha}
J.~M. Cline, K.~Kainulainen, P.~Scott, and C.~Weniger, ``{Update on scalar
  singlet dark matter},''
  \href{http://dx.doi.org/10.1103/PhysRevD.88.055025}{{\em Phys. Rev. D}
  {\bfseries 88} (2013) 055025},
  \href{http://arxiv.org/abs/1306.4710}{{\ttfamily arXiv:1306.4710 [hep-ph]}}.
  [Erratum: Phys.Rev.D 92, 039906 (2015)].

\bibitem{Feng:2014vea}
L.~Feng, S.~Profumo, and L.~Ubaldi, ``{Closing in on singlet scalar dark
  matter: LUX, invisible Higgs decays and gamma-ray lines},''
  \href{http://dx.doi.org/10.1007/JHEP03(2015)045}{{\em JHEP} {\bfseries 03}
  (2015) 045}, \href{http://arxiv.org/abs/1412.1105}{{\ttfamily arXiv:1412.1105
  [hep-ph]}}.

\bibitem{Wu:2016mbe}
H.~Wu and S.~Zheng, ``{Scalar Dark Matter: Real vs Complex},''
  \href{http://dx.doi.org/10.1007/JHEP03(2017)142}{{\em JHEP} {\bfseries 03}
  (2017) 142}, \href{http://arxiv.org/abs/1610.06292}{{\ttfamily
  arXiv:1610.06292 [hep-ph]}}.

\bibitem{GAMBIT:2017gge}
{\bfseries GAMBIT} Collaboration, P.~Athron {\em et~al.}, ``{Status of the
  scalar singlet dark matter model},''
  \href{http://dx.doi.org/10.1140/epjc/s10052-017-5113-1}{{\em Eur. Phys. J. C}
  {\bfseries 77} no.~8, (2017) 568},
  \href{http://arxiv.org/abs/1705.07931}{{\ttfamily arXiv:1705.07931
  [hep-ph]}}.

\bibitem{Casas:2017jjg}
J.~A. Casas, D.~G. Cerde\~no, J.~M. Moreno, and J.~Quilis, ``{Reopening the
  Higgs portal for single scalar dark matter},''
  \href{http://dx.doi.org/10.1007/JHEP05(2017)036}{{\em JHEP} {\bfseries 05}
  (2017) 036}, \href{http://arxiv.org/abs/1701.08134}{{\ttfamily
  arXiv:1701.08134 [hep-ph]}}.

\bibitem{Belyaev:2016lok}
A.~Belyaev, G.~Cacciapaglia, I.~P. Ivanov, F.~Rojas-Abatte, and M.~Thomas,
  ``{Anatomy of the Inert Two Higgs Doublet Model in the light of the LHC and
  non-LHC Dark Matter Searches},''
  \href{http://dx.doi.org/10.1103/PhysRevD.97.035011}{{\em Phys. Rev. D}
  {\bfseries 97} no.~3, (2018) 035011},
  \href{http://arxiv.org/abs/1612.00511}{{\ttfamily arXiv:1612.00511
  [hep-ph]}}.

\bibitem{Mandal:2021yph}
S.~Mandal, R.~Srivastava, and J.~W.~F. Valle, ``{The simplest scoto-seesaw
  model: WIMP dark matter phenomenology and Higgs vacuum stability},''
  \href{http://dx.doi.org/10.1016/j.physletb.2021.136458}{{\em Phys. Lett. B}
  {\bfseries 819} (2021) 136458},
  \href{http://arxiv.org/abs/2104.13401}{{\ttfamily arXiv:2104.13401
  [hep-ph]}}.

\bibitem{Avila:2019hhv}
I.~M. \'Avila, V.~De~Romeri, L.~Duarte, and J.~W.~F. Valle, ``{Phenomenology of
  scotogenic scalar dark matter},''
  \href{http://dx.doi.org/10.1140/epjc/s10052-020-08480-z}{{\em Eur. Phys. J.
  C} {\bfseries 80} no.~10, (2020) 908},
  \href{http://arxiv.org/abs/1910.08422}{{\ttfamily arXiv:1910.08422
  [hep-ph]}}.

\bibitem{Mahanta:2019gfe}
D.~Mahanta and D.~Borah, ``{Fermion dark matter with $N_2$ leptogenesis in
  minimal scotogenic model},''
  \href{http://dx.doi.org/10.1088/1475-7516/2019/11/021}{{\em JCAP} {\bfseries
  11} (2019) 021}, \href{http://arxiv.org/abs/1906.03577}{{\ttfamily
  arXiv:1906.03577 [hep-ph]}}.

\bibitem{Ahriche:2016cio}
A.~Ahriche, K.~L. McDonald, and S.~Nasri, ``{The Scale-Invariant Scotogenic
  Model},'' \href{http://dx.doi.org/10.1007/JHEP06(2016)182}{{\em JHEP}
  {\bfseries 06} (2016) 182}, \href{http://arxiv.org/abs/1604.05569}{{\ttfamily
  arXiv:1604.05569 [hep-ph]}}.

\bibitem{Hagedorn:2018spx}
C.~Hagedorn, J.~Herrero-Garc\'\i{}a, E.~Molinaro, and M.~A. Schmidt,
  ``{Phenomenology of the Generalised Scotogenic Model with Fermionic Dark
  Matter},'' \href{http://dx.doi.org/10.1007/JHEP11(2018)103}{{\em JHEP}
  {\bfseries 11} (2018) 103}, \href{http://arxiv.org/abs/1804.04117}{{\ttfamily
  arXiv:1804.04117 [hep-ph]}}.

\bibitem{Barreiros:2022aqu}
D.~M. Barreiros, H.~B. Camara, and F.~R. Joaquim, ``{Flavour and dark matter in
  a scoto/type-II seesaw model},''
  \href{http://dx.doi.org/10.1007/JHEP08(2022)030}{{\em JHEP} {\bfseries 08}
  (2022) 030}, \href{http://arxiv.org/abs/2204.13605}{{\ttfamily
  arXiv:2204.13605 [hep-ph]}}.

\bibitem{MEGII:2018kmf}
{\bfseries MEG II} Collaboration, A.~M. Baldini {\em et~al.}, ``{The design of
  the MEG II experiment},''
  \href{http://dx.doi.org/10.1140/epjc/s10052-018-5845-6}{{\em Eur. Phys. J. C}
  {\bfseries 78} no.~5, (2018) 380},
  \href{http://arxiv.org/abs/1801.04688}{{\ttfamily arXiv:1801.04688
  [physics.ins-det]}}.

\bibitem{Marciano:1977wx}
W.~J. Marciano and A.~I. Sanda, ``{Exotic Decays of the Muon and Heavy Leptons
  in Gauge Theories},''
  \href{http://dx.doi.org/10.1016/0370-2693(77)90377-X}{{\em Phys. Lett. B}
  {\bfseries 67} (1977) 303--305}.

\bibitem{Cheng:1980tp}
T.~P. Cheng and L.-F. Li, ``{$\mu \to e \gamma$ in Theories With Dirac and
  Majorana Neutrino Mass Terms},''
  \href{http://dx.doi.org/10.1103/PhysRevLett.45.1908}{{\em Phys. Rev. Lett.}
  {\bfseries 45} (1980) 1908}.

\bibitem{Lim:1981kv}
C.~S. Lim and T.~Inami, ``{Lepton Flavor Nonconservation and the Mass
  Generation Mechanism for Neutrinos},''
  \href{http://dx.doi.org/10.1143/PTP.67.1569}{{\em Prog. Theor. Phys.}
  {\bfseries 67} (1982) 1569}.

\bibitem{Langacker:1988up}
P.~Langacker and D.~London, ``{Lepton Number Violation and Massless
  Nonorthogonal Neutrinos},''
  \href{http://dx.doi.org/10.1103/PhysRevD.38.907}{{\em Phys. Rev. D}
  {\bfseries 38} (1988) 907}.

\bibitem{Ilakovac:1994kj}
A.~Ilakovac and A.~Pilaftsis, ``{Flavor violating charged lepton decays in
  seesaw-type models},''
  \href{http://dx.doi.org/10.1016/0550-3213(94)00567-X}{{\em Nucl. Phys. B}
  {\bfseries 437} (1995) 491},
  \href{http://arxiv.org/abs/hep-ph/9403398}{{\ttfamily arXiv:hep-ph/9403398}}.

\bibitem{Toma:2013zsa}
T.~Toma and A.~Vicente, ``{Lepton Flavor Violation in the Scotogenic Model},''
  \href{http://dx.doi.org/10.1007/JHEP01(2014)160}{{\em JHEP} {\bfseries 01}
  (2014) 160}, \href{http://arxiv.org/abs/1312.2840}{{\ttfamily arXiv:1312.2840
  [hep-ph]}}.

\bibitem{Vicente:2014wga}
A.~Vicente and C.~E. Yaguna, ``{Probing the scotogenic model with lepton flavor
  violating processes},'' \href{http://dx.doi.org/10.1007/JHEP02(2015)144}{{\em
  JHEP} {\bfseries 02} (2015) 144},
  \href{http://arxiv.org/abs/1412.2545}{{\ttfamily arXiv:1412.2545 [hep-ph]}}.

\bibitem{BaBar:2009hkt}
{\bfseries BaBar} Collaboration, B.~Aubert {\em et~al.}, ``{Searches for Lepton
  Flavor Violation in the Decays tau+- ---\ensuremath{>} e+- gamma and tau+-
  ---\ensuremath{>} mu+- gamma},''
  \href{http://dx.doi.org/10.1103/PhysRevLett.104.021802}{{\em Phys. Rev.
  Lett.} {\bfseries 104} (2010) 021802},
  \href{http://arxiv.org/abs/0908.2381}{{\ttfamily arXiv:0908.2381 [hep-ex]}}.

\bibitem{Belle-II:2018jsg}
{\bfseries Belle-II} Collaboration, W.~Altmannshofer {\em et~al.}, ``{The Belle
  II Physics Book},'' \href{http://dx.doi.org/10.1093/ptep/ptz106}{{\em PTEP}
  {\bfseries 2019} no.~12, (2019) 123C01},
  \href{http://arxiv.org/abs/1808.10567}{{\ttfamily arXiv:1808.10567
  [hep-ex]}}. [Erratum: PTEP 2020, 029201 (2020)].

\bibitem{Wang:2015saa}
W.~Wang and Z.-L. Han, ``{Radiative linear seesaw model, dark matter, and
  $U(1)_{B-L}$},'' \href{http://dx.doi.org/10.1103/PhysRevD.92.095001}{{\em
  Phys. Rev. D} {\bfseries 92} (2015) 095001},
  \href{http://arxiv.org/abs/1508.00706}{{\ttfamily arXiv:1508.00706
  [hep-ph]}}.

\bibitem{Das:2017ski}
A.~Das, T.~Nomura, H.~Okada, and S.~Roy, ``{Generation of a radiative neutrino
  mass in the linear seesaw framework, charged lepton flavor violation, and
  dark matter},'' \href{http://dx.doi.org/10.1103/PhysRevD.96.075001}{{\em
  Phys. Rev. D} {\bfseries 96} no.~7, (2017) 075001},
  \href{http://arxiv.org/abs/1704.02078}{{\ttfamily arXiv:1704.02078
  [hep-ph]}}.

\bibitem{CarcamoHernandez:2021tlv}
A.~E. C\'arcamo~Hern\'andez, S.~Kovalenko, F.~S. Queiroz, and Y.~S. Villamizar,
  ``{An extended 3-3-1 model with radiative linear seesaw mechanism},''
  \href{http://dx.doi.org/10.1016/j.physletb.2022.137082}{{\em Phys. Lett. B}
  {\bfseries 829} (2022) 137082},
  \href{http://arxiv.org/abs/2105.01731}{{\ttfamily arXiv:2105.01731
  [hep-ph]}}.

\bibitem{Ma:2009gu}
E.~Ma, ``{Radiative inverse seesaw mechanism for nonzero neutrino mass},''
  \href{http://dx.doi.org/10.1103/PhysRevD.80.013013}{{\em Phys. Rev. D}
  {\bfseries 80} (2009) 013013},
  \href{http://arxiv.org/abs/0904.4450}{{\ttfamily arXiv:0904.4450 [hep-ph]}}.

\bibitem{Bazzocchi:2009kc}
F.~Bazzocchi, D.~G. Cerdeno, C.~Munoz, and J.~W.~F. Valle, ``{Calculable
  inverse-seesaw neutrino masses in supersymmetry},''
  \href{http://dx.doi.org/10.1103/PhysRevD.81.051701}{{\em Phys. Rev. D}
  {\bfseries 81} (2010) 051701},
  \href{http://arxiv.org/abs/0907.1262}{{\ttfamily arXiv:0907.1262 [hep-ph]}}.

\bibitem{Baldes:2013eva}
I.~Baldes, N.~F. Bell, K.~Petraki, and R.~R. Volkas, ``{Two radiative inverse
  seesaw models, dark matter, and baryogenesis},''
  \href{http://dx.doi.org/10.1088/1475-7516/2013/07/029}{{\em JCAP} {\bfseries
  07} (2013) 029}, \href{http://arxiv.org/abs/1304.6162}{{\ttfamily
  arXiv:1304.6162 [hep-ph]}}.

\bibitem{CarcamoHernandez:2018hst}
A.~E. C\'arcamo~Hern\'andez, S.~Kovalenko, J.~W.~F. Valle, and C.~A.
  Vaquera-Araujo, ``{Neutrino predictions from a left-right symmetric flavored
  extension of the standard model},''
  \href{http://dx.doi.org/10.1007/JHEP02(2019)065}{{\em JHEP} {\bfseries 02}
  (2019) 065}, \href{http://arxiv.org/abs/1811.03018}{{\ttfamily
  arXiv:1811.03018 [hep-ph]}}.

\bibitem{Mandal:2019oth}
S.~Mandal, N.~Rojas, R.~Srivastava, and J.~W.~F. Valle, ``{Dark matter as the
  origin of neutrino mass in the inverse seesaw mechanism},''
  \href{http://dx.doi.org/10.1016/j.physletb.2021.136609}{{\em Phys. Lett. B}
  {\bfseries 821} (2021) 136609},
  \href{http://arxiv.org/abs/1907.07728}{{\ttfamily arXiv:1907.07728
  [hep-ph]}}.

\bibitem{Casas:2001sr}
J.~A. Casas and A.~Ibarra, ``{Oscillating neutrinos and $\mu \to e, \gamma$},''
  \href{http://dx.doi.org/10.1016/S0550-3213(01)00475-8}{{\em Nucl. Phys. B}
  {\bfseries 618} (2001) 171--204},
  \href{http://arxiv.org/abs/hep-ph/0103065}{{\ttfamily arXiv:hep-ph/0103065}}.

\end{thebibliography}
\end{document}